\documentclass[preprint]{elsarticle}

\usepackage{times} 

\usepackage[all]{xy} 
\def\larrow#1#2#3{\xymatrix{ #3 & #1 \ar[l]_-{#2} }}
\def\longlarrow#1#2#3{\xymatrix{ #3 && #1 \ar[ll]_-{#2} }}

\def\rarrow#1#2#3{\xymatrix{ #1 \ar[r]^-{#2} & #3 }}

\def\verylonglarrow#1#2#3{\xymatrix{ #3 &&& #1 \ar[lll]_-{#2} }}
\usepackage{listings}

\usepackage{amssymb}
\usepackage{MnSymbol}

\usepackage{comment}

\usepackage{algorithmic}
\usepackage{algorithm}

\newcount\td \td=0 
\long \def\Todo#1{\vskip 1ex\noindent\fbox{
\global\advance\td by 1
\textbf{To-do (\number\td):}\\~
\begin{minipage}{.84\textwidth}
#1
\end{minipage}
}\vskip 1ex}
\let\iso=\cong
\def\bang{{!}}
\def\fst{\mathit{p_1}}
\def\snd{\mathit{p_2}}
\def\kr{\mathbin\bigtriangledown}
\def\fork#1#2{#1 \mathbin\bigtriangledown #2}

\def\R{\mathbb{R}}
\def\matlab{\textsc{Matlab}}
\def\fun#1{{\sf #1}}
\def\Matk{\mathit{Mat_K}}           
\def\gje{\mathit{gje}}
\def\msplit#1#2{\left[\begin{array}{c}#1\\\hline#2\end{array}\right]} 
\def\msplitwithvskip#1#2{
\left[\begin{array}{c}#1\rule[-.6em]{0pt}{1.2em}\\\hline#2\rule{0pt}{1.2em}\end{array}\right]} 
\def\meither#1#2{\left[\begin{array}{r|l}#1&#2\end{array}\right]}        
\def\mblock#1#2#3#4{\left[\begin{array}{c|c}#1&#2\\\hline#3&#4\end{array}\right]}
\def\equiv{\Leftrightarrow}
\def\conv#1{#1^\circ}
\def\from{\mathbin\leftarrow}
\def\comp{\mathbin{\cdot}}
\def\just#1#2{\\ &#1& \rule{2em}{0pt} \{ \mbox{\rule[-.7em]{0pt}{1.8em} \small #2 \/} \} \nonumber\\ && }
\def\matrix#1#2#3#4{\begin{bmatrix}#1&#2\\#3&#4\end{bmatrix}}
\def\curry#1{\textbf{curry}\, #1}
\def\vect#1{\textbf{vec}\, #1}              
\def\vectk#1#2{\textbf{vec}_{#1}\, #2}              
\def\unvect#1{\textbf{unvec}\, #1}          
\def\unvectk#1#2{\textbf{unvec}_{#1}\, #2}              
\def\xarrayin#1{\begin{array}{cccccc}#1\end{array}}

\def\MA#1#2#3{#1}
\def\wider#1{~ #1 ~}

\def\eqnnewpage{\end{eqnarray*}\newpage\begin{eqnarray*}&&}
\def\eqnnewpagex{\end{eqnarray}\newpage\begin{eqnarray}&&\nonumber}
\def\implies{\mathbin{\Rightarrow}}
\def\implied{\mathbin{\Leftarrow}}
\def\quotes#1{``#1''}

\newenvironment{mysmatrix}{\left[\begin{smallmatrix}}{\end{smallmatrix}\right]}
\newenvironment{lcbr}{\left\{\begin{array}{l}}{\end{array}\right.}
\def\rcb#1#2#3#4{\def\nothing{}\def\range{#3}\mathopen{\langle}#1 \ #2 \ \ifx\range\nothing::\else: \ #3 :\fi \ #4\mathclose{\rangle}}
\def\p#1{\pi_{#1}}
\def\msplitwithSS#1#2{
\left[\begin{array}{c}#1\\\noalign{\smallskip} \hline \noalign{\smallskip}#2\end{array}\right]}

\def\meitherIL#1#2{\mbox{\footnotesize {$\meither{#1}{#2}$}}}
\def\msplitIL#1#2{\raisebox{0.23ex}{\tiny {$\msplit{#1}{#2}$}}}

\usepackage{color}
\definecolor{blue}{rgb}{0,0,1}
\definecolor{red}{rgb}{1,0,0}
\definecolor{black}{rgb}{0,0,0}

\biboptions{sort}
\newtheorem{thm}{Theorem}

\title{Typing Linear Algebra: A Biproduct-oriented Approach}

\author[uminho]{Hugo Daniel Macedo\corref{cor1}\fnref{fn1}} 
\ead{hmacedo@di.uminho.pt}

\author[uminho]{Jos\'{e} Nuno Oliveira} 
\ead{jno@di.uminho.pt}

\cortext[cor1]{Corresponding author.}
\fntext[fn1]{Partially supported by the \emph{Funda\c{c}\~ao para a Ci\^encia e a
Tecnologia}, Portugal, under grant number {SFRH/BD/33235/2007}} 

\address[uminho]{HASLAB - High Assurance Software Laboratory\\ Universidade do Minho, Braga, Portugal} 

\begin{document}

\begin{abstract}
		Interested in formalizing the generation of fast running code for linear
		algebra applications, the authors show how an index-free, calculational approach
		to matrix algebra can be developed by regarding matrices as morphisms of
		a category with biproducts.
		This shifts the traditional view of matrices as indexed structures to a
		type-level perspective analogous to that of the pointfree algebra of programming.
		The derivation of fusion, cancellation and abide laws from the biproduct
		equations makes it easy to calculate algorithms implementing 
		matrix multiplication, the central operation of matrix algebra,
		ranging from its divide-and-conquer
		version to its vectorization implementation.

		From errant attempts to learn how particular products and coproducts
		emerge from biproducts, not only blocked matrix algebra is
		rediscovered but also a way of extending other operations
		(e.g.\ Gaussian elimination)
		blockwise, in a calculational style, is found.

		The prospect of building biproduct-based type checkers for computer algebra
		systems such as \matlab\texttrademark\ is also considered.
\end{abstract}


\maketitle

~ \hfill
\begin{minipage}{.65\textwidth}\footnotesize\em
``Using matrix notation such a set of simultaneous equations takes the form
$A\comp x=b$ where $x$ is the vector of unknown values, $A$ is the matrix
of coefficients and $b$ is the vector of values on the right side of the
equation. In this way a set of equations has been reduced to a single equation.
This is a tremendous improvement in concision that does not incur any loss
of precision!''
\\ \em
\rule{35ex}{0pt} Roland Backhouse \cite{Ba04a}
\end{minipage}

%
%
\section{Introduction}
In a recent article \cite{DBLP:journals/computer/Parnas10}, David Parnas questions the
traditional use of formal methods in software development,
which he regards unfit
for the software industry.
At the core of Parnas objections lies the contrast
between the current ad-hoc (re)invention of cumbersome mathematical notation,
often a burden to use, and
elegant (thus useful) concepts which are neglected,
often for cultural or (lack of) background reasons.

The question is: {what is it that tells ``good'' and ``bad'' methods apart?}
As Parnas writes, \emph{there is a disturbing gap between software development and traditional engineering disciplines}.
In such disciplines one finds a successful, well-established
mathematical background
essentially made of calculus,
vector spaces, linear algebra and probability theory.
This raises another question: can one hope to share such a successful tradition
in the computing field, or is this definitely a different kind of science,
hostage of formal logics and set theory?

There are signs of change in such direction already,
as interest in the application
of linear algebra techniques to computing seems to be growing, driven by disparate
research interests briefly reviewed below.

Gunther Schmidt, for instance, makes extensive use
of matrix notation, concepts and operations in his recent book on relational mathematics
\cite{Sc10}. This pays tribute to binary relations being just
Boolean matrices.
Of historical relevance, explained in \cite{Mad91}, is the fact of one 
of the first known definitions of relational composition, due to
Charles 
Peirce (1839-1914), being essentially what we understand today as matrix multiplication.

In the area of process semantics, 
Bloom \emph{et al} \cite{BSW96} have developed a categorical,
\emph{machines as matrices} approach to concurrency
\footnote{
Work in this vein can be traced much earlier, back to Conway's work on \emph{regular
algebras} \cite{Co71} and regular algebras of matrices, so elegantly presented
in textbook \cite[Chap.~10]{Ba04a} where the opening quotation of the current paper
is taken from.}; Tr\v{c}ka \cite{Tr09} presents a unifying matrix approach to
the notions of strong, weak and branching bisimulation ranging from labeled
transition systems to Markov reward chains; and Kleene coalgebra is going
quantitative \cite{BSBR11}.

The ``quantum inspiration'' is also pushing computing towards linear algebra foundations.
Focussing on
quantum programming and 
semantics of probabilistic programs, Sernadas \emph{et al} \cite{SRM08b}
adopt linear algebra techniques by regarding probabilistic programs as
linear transformations over suitable vector spaces.
Natural language semantics, too, is going vectorial, as nicely captured by
the aphorism \emph{nouns are vectors, adjectives are matrices} \cite{BZ10}.
In this field of ``quantum linguistics'',
Coecke \emph{et al} \cite{CSC10} have developed a compositional model of meaning
in which the grammatical structure of sentences is expressed in the category
of finite dimensional vector spaces. Unrelated to quantum linguistics but related to
knowledge discovery,
the authors of the current paper show in \cite{MO11-submitted} how to implement data mining
operations
solely based on linear algebra operations. And more examples of the adoption
of linear algebra background in computing could be mentioned.


\section{Typing Linear Algebra}

One R\&D field whose core lies in linear algebra (LA)
is the automatic generation of fast running code for LA applications
running on parallel architectures \cite{Johnson:90,Pueschel:05,TSoPMC,Franchetti:09}.
The sophisticated techniques developed in this direction of research
call for matrix multiplication as kernel operator, whereby matrices
are viewed and transformed in an index-free way \cite{Franchetti:09}.

Interestingly, the successful
language SPL \cite{Pueschel:05}
used in generating automatic parallel code
has been created envisaging the same
principles as advocated by the purist computer scientist:
index-free abstraction 
and composition (multiplication) as a kernel way of connecting objects
of interest (matrices, programs, etc).

There are several domain specific languages (DSLs) bearing such purpose in
mind \cite{Pueschel:05,TSoPMC,Franchetti:09}. However, they arise
as programming dialects with 
poor type checking. 
Of popular use and suffering from the same weakness one finds
the widespread \matlab~\footnote{
\matlab\ \texttrademark\ is a trademark of The MathWorks \textregistered.
}
library of matrix operations, in which users have to keep track of dimensions
all the way through and raise exceptions wherever ``expressions don't fit
with each other''.
This 
hinders effective use of such languages and libraries,
calling for a ``type structure'' in linear algebra systems similar to
that underlying modern functional programming languages such as Haskell,
for instance \cite{Ha03}.

It so happens that, in the same way function composition is the kernel operation
of functional programming, leading to the \emph{algebra of programming} \cite{BM97},
so does matrix multiplication once matrices are viewed and transformed in an
index-free way. Therefore, rather than interpreting the product $A B$ of matrices $A$ and $B$
as an algorithm for computing a new matrix $C$ out of $A$ and $B$, and trying
to build and explain matrix algebra systems out of such an algorithm, one wishes
to abstract from \emph{how} the operation is carried out.
Instead, the emphasis is put on its
type structure, regarded as the pipeline $A \comp B$
(to be read as ``A after B''), as if $A$ and $B$ were functions
\begin{eqnarray}
	C& =&A \comp B
\end{eqnarray}
or binary relations --- the actual building block of the algebra of programming
\cite{BM97}. In this discipline, relations are viewed as (typed) composable
arrows (morphisms) which can be combined in a number of ways, namely by joining
or intersecting relations of the same type, reversing them (thus
swapping their source and target types), and so on.

If relations, which are Boolean matrices, can be regarded as morphisms of a
suitable mathematical framework \cite{FS90,BM97}, why not regard arbitrary matrices in the same way?
%
This matches with the categorical characterization of matrices,
which can be traced back to Mac Lane \cite{MacLane98:CategoriesWM}, whereby
matrices are regarded as arrows in a category whose objects are natural numbers
(matrix dimensions):
\begin{eqnarray}
	A = 
	\begin{bmatrix}
		a_{11} & \ldots & a_{1n}\\
		\vdots & \ddots & \vdots\\
		a_{m1} & \ldots & a_{mn} 
	\end{bmatrix}
		_{m \times n}
&
	\rule{.1\textwidth}{0pt}
&
	\larrow{n}{A}{m}
	\label{eq:091205b}
\end{eqnarray}

Such a category $\Matk$ of matrices over a field $K$
merges categorical products and coproducts into a single construction termed
\emph{biproduct} \cite{MacLane98:CategoriesWM}. Careful analysis of the biproduct
axioms as a system of equations provides a rich \emph{palette} of
constructs for building matrices from smaller ones. In \cite{MO10a}
we developed an approach to matrix blocked operation stemming
from one particular solution to such equations, which in fact
offers explicit operators for building block-wise
matrices (row and column-wise) as defined by \cite{BSW96}.
We also showed
how divide-and-conquer algorithms for linear algebra arise from biproduct laws
emerging from the underlying categorial basis.

In the current paper we elaborate on \cite{MO10a} and show how biproduct-orientation
leads into a simple, polymorphic type system for linear algebra.
In the same way the categorial approach to functional programming
--- types-as-objects, functions-as-morphisms, 
etc \cite{BM97}
--- leads into a widely acclaimed type-system,
so one expects categories of matrices to offer a basis for typing linear algebra,
as will be shown in this paper.
Resistance to adopting such a categorial, but simple type system
entails the need for more elaborate type mechanisms such as eg.\ \emph{dependent
types} \cite{BD09}~\footnote{In fact,
typing matrix operators provides a popular illustration of dependent types \cite{BD09}.}.

The paper includes three illustrations of biproduct-orientation:
the implementation of matrix-matrix multiplication (MMM),
a blocked version of the Gauss-Jordan elimination algorithm
and a thorough study of vectorization, required in mapping matrices into
computers' linear storage.
Altogether,
the paper gives the details of a constructive approach
to matrix algebra operations leading to elegant, index-free proofs of the corresponding
algorithms.

\paragraph{Structure of the paper}
The remainder of this paper is structured as follows.
Section \ref{sec:110107a} introduces the reader to categories of matrices and biproducts.
Section \ref{sec:chasing} finds solutions to the biproduct equations, in
particular those which explain blocked-matrix operations.
Sections \ref{sec:110107b} and \ref{sec:110107c} develop a calculational approach to
blocked linear algebra and present an application --- that of calculating the
nested-loop implementation of MMM.
Section \ref{sec:110107d} shows how to develop biproduct algebra for applications,
illustrated by the synthesis of a blocked-version of Gauss-Jordan elimination.
Sections \ref{sec:101218b} and \ref{sec:110107e} show how
the algebra of matrix vectorization emerges from a self-adjunction in the category
of matrices whose unit and counit are expressed in terms of the underlying biproduct.
Section \ref{sec:101229d} shows how to refine linear algebra operators once matrices are
represented by vectors.

The remaining sections review related work and conclude, giving pointers for future research.

%
%
\section{The Category of Matrices \textbf{Mat$_K$}}
\label{sec:110107a}

Matrices are mathematical objects that can be traced back to ancient times,
documented as early as 200 BC \cite{allenby}. The word ``matrix'' was introduced
in the western culture much later, in the 1840's, by the mathematician
James Sylvester (1814-1897)
when both matrix theory and linear algebra emerged.

The traditional way of viewing matrices as rectangular tables (\ref{eq:091205b})
of elements or entries (the ``container view'') which in turn are other
mathematical
objects such as e.g.\ complex numbers (in general: inhabitants of the field $K$
which underlies $\Matk$), encompasses as special cases one column
and one line matrices, referred to as column (resp.\ row) \emph{vectors},
that is, matrices of shapes
\begin{eqnarray*}
	v = \begin{bmatrix}  
	v_{1} \\
	\vdots\\
	v_{m}
	\end{bmatrix}
&
	\mbox{ ~~~~and~~~~}
&
	w = \begin{bmatrix}  
	w_{1} & \ldots & w_{n}\\
	\end{bmatrix}
\end{eqnarray*}

\paragraph{What is a matrix?}
The standard answer to this question is to regard matrix $A$ (\ref{eq:091205b})
as a computation unit, or transformation,
which commits itself to producing a (column) vector of size $m$  provided it 
is supplied with a (column) vector of size $n$.
How is such output produced?
Let us abstract from this at this stage and look at diagram
\begin{eqnarray*}
\xymatrix{
	m
&
	n
		\ar[l]_{A}
&
	1
		\ar[l]_{v}
		\ar@/^1pc/[ll]^{w}
}
\end{eqnarray*}
arising from depicting the situation above in arrow notation.
This suggests a pictorial representation of the product of
matrix $A_{m\times n}$ and matrix $B_{n \times q}$,
yielding a new matrix $C = (A B)_{m \times q}$ 
with dimensions $m \times q$, as follows,
\begin{eqnarray}
	\xymatrix{ m   & n\ar[l]_{A}  & q \ar[l]_{B} \ar@/^1pc/[ll]^{C = A \comp B} }
	\label{eq:comp}
\end{eqnarray}
which automatically ``type-checks'' the construction: the ``target'' of 
\(
	\larrow q B n
\)
simply matches the ``source'' of 
\(
	\larrow n A m
\)
yielding a matrix whose type
\(
	\larrow q \relax m
\)
is the composition of the given types.


Having defined matrices as composable arrows in a category, we need to define
its identities \cite{MacLane98:CategoriesWM}:
for every object $n$, there must be an arrow of type
\(
	\larrow n ~ n
\)
which is the unit of composition. This is nothing but the identity matrix of size $n$,
which will be denoted by
\(
	\larrow n {id_n} n
\)
or
\(
	\larrow n {1} n
\),
indistinguishably.
Therefore,
for every matrix $\larrow n A m$, equalities
\begin{eqnarray}
	id_m \comp A \wider= A \wider= A \comp id_n
&
	~~~~~~~~~
&
\xymatrix{
        n
		\ar[d]_-{A}
&
        n
		\ar[l]_-{id_n}
		\ar[d]^-{A}
		\ar[dl]^-{A}
\\
        m
&
        m
		\ar[l]^-{id_m}
}
\label{eq:natid}
\end{eqnarray}
hold.
(Subscripts $m$ and $n$ can be omitted wherever the underlying diagrams are assumed.)

\paragraph{Transposed matrices} 
One of the kernel operations of linear algebra is \emph{transposition}, whereby
a given matrix changes shape by turning its rows into columns and vice-versa.
Type-wise, this means converting an arrow $\larrow m A n$ into an arrow $\larrow n {A^\top} m$,
that is, source and target types (dimensions) switch over.
By analogy with relation algebra, where a similar operation is termed \emph{converse}
and denoted $\conv A$, we will use this notation instead of $A^\top$
and will say ``$A$ converse'' wherever reading $\conv A$.
Index-wise, we have, for $A$ as in (\ref{eq:091205b}):
\begin{eqnarray*}
\conv{A} = \begin{bmatrix}
a_{11} & \ldots & a_{m1}\\
\vdots & \ddots & \vdots\\
a_{1n} & \ldots & a_{mn}
\end{bmatrix}    
&
	\rule{.1\textwidth}{0pt}
&
	\larrow{n}{\conv A}{m}
\end{eqnarray*}
Instead of telling how transposition is carried out index-wise,
again we prefer to stress on (index-free) properties of this operation such
as,
among others,
idempotence and contravariance:
\begin{eqnarray}
	\conv{(\conv A)} & = & A
\\
	\conv{(A \comp B)} &=& \conv B \comp \conv A
	\label{eq:091206b}
\end{eqnarray}

\paragraph{Bilinearity}
Given two matrices of the same type $\larrow n {A,B} m$
(i.e., in the same homset of $\Matk$) it makes sense to add them
up index-wise, leading to matrix $A+B$ where symbol $+$ promotes
the underlying element-level additive operator to matrix-level.
Likewise, additive unit element $0$ is promoted to matrix $0$ wholy filled with $0$s,
the unit of matrix addition and zero of matrix composition:
\begin{eqnarray}
&
\xarrayin{
	A + 0 & = & A & = & 0 + A
} &
	\label{eq:111008a}
\\
&
\xarrayin{
	A \comp 0 & = & 0 & = & 0 \comp  A
}&
\end{eqnarray}
In fact, matrices form an \emph{Abelian category}: each homset in the category
forms an additive Abelian (i.e.\ commutative) group with respect to which composition is bilinear:
\begin{eqnarray}
	A\comp (B + C) &=& A\comp B + A\comp C
	\label{eq:090403b}
\\
	(B + C)\comp A &=& B\comp A + C\comp A
	\label{eq:091205a}
\end{eqnarray}
Polynomial expressions (such as in the properties above) denoting matrices
built up in an index-free way from addition and composition play a major role in matrix
algebra. This can be appreciated in the explanation of the very important concept of
a \emph{biproduct} \cite{MacLane98:CategoriesWM,MacLane99:Algebra} which follows.

\paragraph{Biproducts}  In an Abelian category,
a \emph{biproduct} diagram for the objects $m, n$ is a diagram of shape

\begin{displaymath}
\xymatrix{
m \ar@<-1ex>[r]_{i_1}  &   r \ar@<-1ex>[l]_{\pi_1} \ar@<1ex>[r]^{\pi_2}   & n \ar@<1ex>[l]^{i_2}\\
}
\end{displaymath}
whose arrows $\pi_1$, $\pi_2$, $i_1$, $i_2$ satisfy the identities which follow:
\begin{eqnarray}
	\pi_1 \comp i_1 &=& id_m \label{eq:biprod:a}
\\
	\pi_2 \comp i_2 &=& id_n \label{eq:biprod:b}
\\
	i_1 \comp \pi_1 + i_2 \comp \pi_2 &=& id_{r}
	\label{eq:biprod:c}
\end{eqnarray}
Morphisms $\p i$ and $i_i$ are termed \emph{projections} and
\emph{injections}, respectively. From the underlying arithmetics
one easily derives the following orthogonality properties (details in
the appendix):
\begin{eqnarray}
	\pi_1 \comp i_2 = 0
	\label{eq:111006a}
\\
	\pi_2 \comp i_1 = 0
	\label{eq:091211a}
\end{eqnarray}

One wonders: how do biproducts relate to products and co-products in the category?
The answer in Mac Lane's \cite{MacLane98:CategoriesWM} words is as follows:

\begin{quote}\em\small
Theorem 2: Two objects $a$ and $b$ in Abelian category $A$ have a product
in $A$ iff they have a biproduct in $A$. Specifically, given a biproduct
diagram, the object $r$ with the projections $\pi_1$ and $\pi_2$ is a product
of $m$ and $n$, while, dually, $r$ with $i_1$ and $i_2$ is a coproduct. In
particular, two objects $m$ and $n$ have a product in $A$ if and only if
they have a coproduct in $A$.
\end{quote}
The diagram and definitions below depict how products and coproducts arise
from biproducts (the product diagram is in the lower half; the upper half
is the coproduct one):

\noindent\begin{minipage}{0.5\linewidth}
	\[
	\xymatrix@R+1.1ex{
	 & m  &  \\
	\ar[ur]^{A} n 
	\ar@<-.7ex>[r]_{i_1}  &   n+p 
	\ar[u]|(.45)*+{\mbox{\tiny$\meither{A}{B}$}} 
	\ar@<-.7ex>[l]_{\p1} 
	\ar@<.7ex>[r]^{\p2}  & p 
	\ar@<.7ex>[l]^{i_2} 
	\ar[ul]_{B} \\ 
	 &
	\ar[ul]^{C} t  
	\ar[u]|(.44)*+<2pt>{\mbox{\tiny$\msplit{C}{D}$}} 
	\ar[ur]_{D} &  \\
	}
	\]
\end{minipage}
\begin{minipage}{0.5\linewidth}
\begin{eqnarray}
	\meither{A}{B} &=& A \comp \p1 + B \comp \p2 \label{eq:meither:def}\\
\nonumber\\
	\msplit{C}{D} &=& i_1 \comp C  + i_2 \comp D \label{eq:msplit:def}
\end{eqnarray}
\end{minipage}

By analogy with the algebra of programming \cite{BM97},
expressions \meitherIL{A}{B} and \msplitIL{C}{D} will be read
``$A$ junc $B$'' and ``$C$ split $D$'', respectively.
What is the intuition behind these combinators, which come out of the blue
in texts such as e.g.\ \cite{BSW96}?
Let us start by a simple illustration, for $m=n=2$, $p=1$,
$A = \begin{mysmatrix} 1 & 2\\4 & 5\end{mysmatrix}$,  $B = \begin{mysmatrix} 3 \\ 6\end{mysmatrix}$, $\p1 = \begin{mysmatrix} 1 & 0 & 0\\ 0 & 1 & 0\end{mysmatrix}$ and $\p2 = \begin{mysmatrix}
0 & 0 & 1\end{mysmatrix}$.
Then (\ref{eq:meither:def}) instantiates as follows:
\begin{eqnarray*}
&&
	\meither{A}{B}
=
	A \comp \p1 + B \comp \p2
\just={ instantiation }
	\meither{\begin{bmatrix} 1 & 2\\4 & 5\end{bmatrix}}
		{\begin{bmatrix} 3 \\ 6\end{bmatrix}}
=
	\begin{bmatrix} 1 & 2\\4 & 5\end{bmatrix} \comp  
		\begin{bmatrix} 1 & 0 & 0\\ 0 & 1 & 0\end{bmatrix} 
	+  
	\begin{bmatrix} 3 \\ 6\end{bmatrix} \comp 
		\begin{bmatrix} 0 & 0 & 1\end{bmatrix} 
\just={ composition (\ref{eq:comp})}
	\meither{\begin{bmatrix} 1 & 2\\4 & 5\end{bmatrix}}
		{\begin{bmatrix} 3 \\ 6\end{bmatrix}}
=
	\begin{bmatrix} 1 & 2 & 0 \\ 4 & 5 & 0\end{bmatrix} 
	+  
	\begin{bmatrix} 0 & 0 & 3 \\ 0 & 0 & 6\end{bmatrix} 
\just={ matrix addition (\ref{eq:111008a}) }
	\meither{\begin{bmatrix} 1 & 2\\4 & 5\end{bmatrix}}
		{\begin{bmatrix} 3 \\ 6\end{bmatrix}}
=
	\begin{bmatrix} 1 & 2 & 3 \\ 4 & 5 & 6\end{bmatrix} 
\end{eqnarray*}
A similar exercise would illustrate the split combinator (consider eg.\ transposing 
all arrows).

Expressed in terms of definitions (\ref{eq:meither:def})
and (\ref{eq:msplit:def}), axiom (\ref{eq:biprod:c}) rewrites to both
\begin{eqnarray}
	\meither{i_1}{i_2} & = & id
	\label{eq:091028a}
\\
	\msplit{\pi_1}{\pi_2} & = & id
	\label{eq:091028b}
\end{eqnarray}
somehow suggesting that the two injections and the two projections ``decompose''
the identity matrix. On the other hand, each of (\ref{eq:091028a},\ref{eq:091028b})
has the shape of a \emph{reflection} corollary \cite{BM97} of some universal property.
Below we derive such a property for \meitherIL{A}{B},
\begin{eqnarray}
	X = \meither{A}{B}
& \equiv &
	\begin{lcbr}
	X \comp i_1 = A
\\
	X \comp i_2 = B
	\end{lcbr}
	\label{eq:091206c}
\end{eqnarray}
from the underlying biproduct equations, by two-way implication:
\begin{eqnarray*}
&&
	X = \meither{A}{B}
\just\equiv{ identity (\ref{eq:natid}) ; (\ref{eq:meither:def}) }
	X \comp id = A \comp \pi_1 + B \comp \pi_2
\just\equiv{ (\ref{eq:biprod:c}) }
	X \comp (i_1 \comp \p1 + i_2 \comp \p2) = A \comp \pi_1 + B \comp \pi_2
\just\equiv{ bilinearity (\ref{eq:090403b}) }
	X \comp i_1 \comp \p1 + X \comp i_2 \comp \p2 = A \comp \pi_1 + B \comp \pi_2
\just\implies{ Leibniz (twice) }
	\begin{lcbr}
	(X \comp i_1 \comp \p1 + X \comp i_2 \comp \p2) \comp i_1 = (A \comp \pi_1 + B \comp \pi_2) \comp i_1
\\
	(X \comp i_1 \comp \p1 + X \comp i_2 \comp \p2) \comp i_2 = (A \comp \pi_1 + B \comp \pi_2) \comp i_2
	\end{lcbr}
%
%
\just\equiv{ bilinearity (\ref{eq:091205a}) ; biproduct (\ref{eq:biprod:a},\ref{eq:biprod:b}) ; orthogonality (\ref{eq:091211a}) }
	\begin{lcbr}
	X \comp i_1 + X \comp i_2 \comp 0 = A  + B \comp 0
\\
	X \comp i_1 \comp 0 + X \comp i_2 = A \comp 0 + B
	\end{lcbr}
\just\equiv{ trivial }
	\begin{lcbr}
	X \comp i_1 = A
\\
	X \comp i_2 = B
	\end{lcbr}
\just\implies{ Leibniz (twice) }
	\begin{lcbr}
	X \comp i_1 \comp \p1 = A \comp \pi_1
\\
	X \comp i_2 \comp \p2 = B \comp \pi_2
	\end{lcbr}
\just\implies{ Leibniz }
	X \comp i_1 \comp \p1 + X \comp i_2 \comp \p2 = A \comp \pi_1 + B \comp \pi_2
\just\equiv{ as shown above }
	X = \meither{A}{B}
\end{eqnarray*}
The derivation of the universal property of \msplitIL{C}{D},
\begin{eqnarray}
X = \msplit C D & \equiv &
        \left\{
                \begin{array}{rcl}
                        \pi_1 \comp X = C
                \\
                        \pi_2 \comp X = D
                \end{array}
        \right.
	\label{eq:091206d}
\end{eqnarray}
is (dually) analogous.

Last but not least, we stress that injections and projections in a biproduct
are unique. Thus, for instance,
\begin{eqnarray}
                        A \comp \msplit C D = C
                \wider\land
                        B \comp \msplit C D = D
	& \wider\equiv &
	A = \p1 \land B = \p2
	\label{eq:101221h}
\end{eqnarray}
holds~\footnote{Easy to check:
from right to left, just let $X := \msplitIL{C}{D}$ in (\ref{eq:101221h})
and simplify; in the opposite direction, let $C,D := A,B$ in (\ref{eq:101221h})
and note that $\msplitIL{A}{B} =id$ due to \emph{split} uniqueness.
}.

\paragraph{Remarks concerning notation}
Outfix notation such as that used in \emph{splits} and \emph{juncs}
provides for unambiguous parsing of matrix algebra expressions.
Concerning infix operators (such as eg.\ composition, $+$) and unary ones
(eg.\ converse, and others to appear) the following conventions will be
adopted for saving parentheses:
(a) unary and prefix operators bind tighter than binary;
(b) multiplicative binary operators bind tighter than additive ones;
(c) matrix multiplication (composition) binds tighter than any other multiplicative
operator (eg.\ Kronecker product, to appear later).

We will resort to \matlab\ notation to illustrate the main constructions of the paper.
For instance, \emph{split} \msplitIL{A}{B} (resp.\ \emph{junc} \meitherIL{A}{B})
is written as \texttt{[A ; B]} (resp.\ \texttt{\mbox{[A B]}}) in \matlab.
More elaborate constructs will be encoded in the form of \matlab\ functions.

\paragraph{Parallel with relation algebra} 
Similar to matrix algebra, relation algebra \cite{TG87,BM97,Sc10} can also be explained 
in terms of biproducts once morphism addition (\ref{eq:biprod:c}) is interpreted as
relational union, object union as disjoint union, $i_1$ and $i_2$ as the corresponding
injections and $\p1$, $\p2$ their converses, respectively~\footnote{%
Note that orthogonality (\ref{eq:111006a}, \ref{eq:091211a}) is granted by the disjoint union construction itself.}.
Relational product should not, however, be confused with the \emph{fork} 
construct \cite{Fr02} in fork (relation) algebra, which involves pairing.
(For this to become a product one has to restrict to functions.)

It is worth mentioning that the matrix approach to relations,
as intensively stressed in \cite{Sc10}, 
is not restricted to set-theoretic models of allegories. For instance,
Winter \cite{Wi00} builds categories of matrices on top of
categories of relations.

In the next section we show that the converse relationship (duality) between
projections and injections is not a privilege of relation algebra: the most
intuitive biproduct solution in the category of matrices also offers such
a duality.

\section{Chasing biproducts}\label{sec:chasing}
Let us now address the intuition behind products and coproducts of matrices.
This has mainly to do with the interpretation of projections $\pi_1$, $\pi_2$
and injections $i_1$, $i_2$ arising as solutions of biproduct
equations (\ref{eq:biprod:a},\ref{eq:biprod:b},\ref{eq:biprod:c}).
Concerning this,
Mac~Lane \cite{MacLane98:CategoriesWM} laconically writes:
\begin{quotation}\em\small\noindent
	``In other words, the [biproduct] equations contain the familiar calculus of
	matrices.''
\end{quotation}
In what way? The answer to this question proves more interesting than it seems at first,
because of the multiple solutions arising from a non-linear
system of three equations (\ref{eq:biprod:a},\ref{eq:biprod:b},\ref{eq:biprod:c}) with four variables.
%
In trying to exploit this freedom we became aware that each solution
offers a particular way of putting matrices together via the
corresponding ``junc'' and ``split'' combinators.

Our inspection of solutions started by reducing the ``size'' of the objects
involved and experimenting with the smaller biproduct depicted below:

\begin{displaymath}
	\xymatrix{ 1 \ar@<-.5ex>[r]_{i_1} & 1+1 \ar@<-.5ex>[l]_{\pi_1} \ar@<.5ex>[r]^{\pi_2} & 1 \ar@<.5ex>[l]^{i_2}
	} 
\end{displaymath}

\noindent The ``puzzle'' in this case is more manageable,

\[ \left\{
\begin{array}{lcl}
	\pi_1 \comp i_1 & = & [1]\\
	\pi_2 \comp i_2 & = & [1]\\
	i_1 \comp \pi_1 + i_2 \comp \pi_2 & = & \begin{bmatrix}	1 & 0\\ 0 & 1 \end{bmatrix} \\
\end{array}
\right. \]

\noindent yet the set of solutions is not small. We used the Mathematica
software \cite{wolfram1988mathematica} to solve this system by inputting the projections and
injections as suitably typed matrices leading to a larger, non-linear system:

\[ \left\{
\begin{array}{lcl}
    \left[
    \begin{array}{cc}
     \pi_{11} & \pi_{12}
    \end{array}
    \right] \comp \left[
    \begin{array}{c}
     i_{11} \\
     i_{12}
    \end{array}
    \right] & = & [1]
\rule[-4ex]{0pt}{1pt}
\\
    \left[
    \begin{array}{cc}
     \pi_{21} & \pi_{22}
    \end{array}
    \right] \comp \left[
    \begin{array}{c}
     i_{21} \\
     i_{22}
    \end{array}
    \right] & = & [1]
\\
    \left[
    \begin{array}{c}
     i_{11} \\
     i_{12}
    \end{array}
    \right] \comp \left[
    \begin{array}{cc}
     \pi_{11} & \pi_{12}
    \end{array}
    \right] + \left[
    \begin{array}{c}
     i_{21} \\
     i_{22}
    \end{array}
    \right] \comp \left[
    \begin{array}{cc}
     \pi_{21} & \pi_{22}
    \end{array}
    \right] & = & \begin{bmatrix}    1 & 0\\ 0 & 1 \end{bmatrix} \\
\end{array}
\right. \]
This was solved using the standard \text{Solve} command obtaining the output
presented in Figure \ref{tab:Mathematica}, which offers several solutions.
Among these we first picked the one which purports the most intuitive reading
of the \emph{junc} and \emph{split} combinators --- that of simply gluing
matrices vertically and horizontally (respectively) with no further computation
of matrix entries:

\begin{figure}[t]
\begin{center}	
\(
\begin{array}{c}
\pmb{\text{sol}=\text{Simplify}[\text{Solve}[\{\text{pi1}.\text{i1}==\text{I1},\text{pi2}.\text{i2}==\text{I1},\text{i1}.\text{pi1}+\text{i2}.\text{pi2}==\text{I2}\}]]}
\\[0.3em]
\text{Solve}\text{::}\text{svars}: \text{Equations may not give solutions for all $\texttt{"}$solve$\texttt{"}$ variables. }
\\[0.3em]
\left\{\left\{i_{11}\to -\dfrac{\pi_{22}}{\pi_{12} \pi_{21}},i_{12}\to \dfrac{1}{\pi_{12}},i_{21}\to
\dfrac{1}{\pi_{21}},i_{22}\to 0,\pi_{11}\to 0\right\},\right.
\\
\left.\left\{i_{11}\to \dfrac{\pi_{22}}{-\pi_{12} \pi_{21}+\pi_{11}
\pi_{22}},i_{12}\to \dfrac{\pi_{21}}{\pi_{12} \pi_{21}-\pi_{11} \pi_{22}},
\right. \right.
\\
\qquad \left.\left. i_{21}\to \dfrac{\pi_{12}}{\pi_{12}
\pi_{21}-\pi_{11} \pi_{22}},i_{22}\to \dfrac{\pi_{11}}{-\pi_{12} \pi_{21}+\pi_{11} \pi_{22}}\right\} \right\}
\end{array}
\)
\end{center}
\caption{Fragment of Mathematica script}\label{tab:Mathematica}
\end{figure}

\begin{eqnarray*}
&&
\pi_1 = \begin{bmatrix}1 & 0\end{bmatrix} \qquad \pi_2 = \begin{bmatrix}0 & 1\end{bmatrix}
\\
&&
i_1 = \begin{bmatrix} 1 \\ 0\end{bmatrix} \qquad \qquad i_2 = \begin{bmatrix}0 \\ 1\end{bmatrix}
\end{eqnarray*}

\noindent
Interpreted in this way, \msplitIL{A}{B} (\ref{eq:msplit:def}) and \meitherIL{A}{B}
(\ref{eq:meither:def}) are the block gluing matrix operators which one can
find in \cite{BSW96}. Our choice of notation --- $A$ above $B$ in the case
of  (\ref{eq:msplit:def}) and $A$ besides $B$ in the case of  (\ref{eq:meither:def})
reflects this semantics. 


The obvious generalization of this solution to higher dimensions of the problem leads
to the following matrices with identities of size $m$ and $n$ in the 
appropriate place, so as to properly typecheck~\footnote{
Projections $\p1,\p2$ (resp.\ injections $i_1,i_2$) are referred to as \emph{gather}
(resp.\ \emph{scatter})
matrices in \cite{Voronenko:08}. 
\matlab's (untyped) notation for projection  $\p1$ and
injection $i_1$ in (\ref{eq:101221a}) is 
\texttt{eye(m,m+n)} and \texttt{eye(m+n,m)}, respectively.
Consistently, \texttt{eye(n,n)} denotes $id_n$.
Matrices $\p2$ and $\p2$ can be programmed using \matlab's \texttt{eye} and \texttt{zeros} ---
see Listing \ref{lst:101224a} further on.}: 
\begin{eqnarray}
\begin{array}{lll}
	\pi_1 = \longlarrow{m+n}{\meither{id_m ~}{~0}}{m}
& ~~~,~~~ &
	\pi_2 = \longlarrow{m+n}{\meither{0 ~}{~id_n}}{n}
\\
	i_1 = \longlarrow{m}{\msplit{id_m}{0}}{m+n}
& ~~~,~~~ &
	i_2 = \longlarrow{n}{\msplit0{id_m}}{m+n}
\end{array}
	\label{eq:101221a}
\end{eqnarray}
The following diagram pictures not only the construction of this biproduct
but also the biproduct (\ref{eq:biprod:a},\ref{eq:biprod:b}) and orthogonality
(\ref{eq:111006a}, \ref{eq:091211a}) equations
--- check the commuting triangles:
\begin{eqnarray*}
\xymatrix@R-1ex@C-1ex{
&
	m
\\
	m
		\ar[dr]_{0}
		\ar[ur]^{id_m}
		\ar[r]_{i_1}
&
	m+n
		\ar[u]|(.45)*+{\mbox{\scriptsize $\p1$}}
		\ar[d]|(.45)*+{\mbox{\scriptsize $\p2$}}
&
	n
		\ar[l]^{i_2}
		\ar[ul]_{0}
		\ar[dl]^{id_n}
\\
&
	n 
&
}
\end{eqnarray*}
By inspection, one immediately infers the same duality found in relation algebra,
\begin{eqnarray}
	\conv{\p1} = {i_1} & , & \conv{\p2} = {i_2}
	\label{eq:091206a}
\end{eqnarray}
whereby \emph{junc} (\ref{eq:meither:def}) and \emph{split} (\ref{eq:msplit:def})
become self dual:
\begin{eqnarray}
&&
	\conv{\meither{R}{S}}
	\nonumber
\just={ (\ref{eq:meither:def}) ; (\ref{eq:091206b}) }
	\conv{\p1} \comp \conv{R} + \conv{\p2} \comp \conv{S}
	\nonumber
\eqnnewpagex
\just={ (\ref{eq:091206a}) ; (\ref{eq:msplit:def}) }
	\msplit{\conv R}{\conv S}
	\label{eq:110103e}
\end{eqnarray}

This particular solution to the biproduct equations captures what in the
literature is meant by \emph{blocked} matrix algebra, a generalization of the standard
element-wise operations to sub-matrices, or blocks, leading to
\emph{divide-and-conquer}
versions of the corresponding algorithms.
The next section shows the exercise of deriving such laws, thanks to 
the algebra which emerges from the universal properties of the
block-gluing matrix combinators
\emph{junc} (\ref{eq:091206c})
and \emph{split} (\ref{eq:091206d}).
We combine the standard terminology with that borrowed from the algebra of
programming \cite{BM97} to stress the synergy between blocked matrix algebra
and relation algebra.


%
%
\section{Blocked Linear Algebra --- calculationally!}
\label{sec:110107b}
Further to {reflection laws} (\ref{eq:091028a},\ref{eq:091028b}),
the derivation of the following equalities from universal properties 
(\ref{eq:091206c},\ref{eq:091206d}) is a standard exercise in (high) school
algebra, where capital letters $A$, $B$, etc.\ denote suitably typed matrices
(the types, i.e.\ dimensions, involved in each equality can be inferred by drawing the corresponding diagram):
\begin{itemize}
\item	Two ``fusion''-laws:
\begin{eqnarray}
	C\comp \meither A B &=& \meither{C\comp A}{C\comp B} \label{eq:meither:fusion}
\\
	\msplit A B \comp C &=& \msplit{A\comp C}{B\comp C} \label{eq:msplit:fusion}
\end{eqnarray}
\item	Four ``cancellation''-laws~\footnote{
Recall (\ref{eq:101221h}).}:
\begin{eqnarray}
	\meither A B \comp i_1 =  A
& , &
	\meither A B \comp i_2 =  B
	\label{eq:091206e}
\\
	\p1 \comp \msplit  A B =  A
& , &
	\p2 \comp \msplit  A B =  B
	\label{eq:111007b}
\end{eqnarray}
\item	Three ``abide''-laws~\footnote{Neologism ``abide'' (= ``above and beside'')
was introduced by Richard Bird \cite{Bi89} as a generic name for algebraic laws
in which two binary operators written in infix form change place between
\quotes{above} and \quotes{beside}, e.g.\
\begin{eqnarray*}
	\frac a b \times \frac c d & = & \frac {a \times c}{b \times d}
\end{eqnarray*}
 }: the \emph{junc/split} {exchange law}
\begin{eqnarray}
	\xarrayin{
	\msplitwithvskip{\meither A B}{\meither C D} & = & \meither{\msplit A C}{\msplit B D}
	& = & \mblock A B C D
	}
	\label{eq:exc}
\end{eqnarray}
which tells the equivalence between row-major and column-major construction
of matrices (thus the four entry \emph{block} notation on the right),
and two \emph{blocked addition} laws:
\begin{eqnarray}
	\meither A B + \meither C D &=& \meither{A+C}{B+D}
	\label{eq:abide:+}
\\
	\msplit A B + \msplit C D &=& \msplit{A+C}{B+D}	
	\label{eq:abide:x}
\end{eqnarray}
\item	Two structural equality laws (over the same biproduct):
\begin{eqnarray}
	\meither A B = \meither C D & \equiv & A = C \land B = D
	\label{eq:101221g}
\\
	\msplit A B = \msplit C D & \equiv & A = C \land B = D
	\label{eq:110103f}
\end{eqnarray}
\end{itemize}

The laws above are more than enough for us to derive standard linear algebra
rules and algorithms in a calculational way.
As an example of their application we provide a simple proof of the rule
which underlies \emph{divide-and-conquer} matrix multiplication:
\begin{eqnarray}
	\meither A B\comp \msplit C D & = & A\comp C + B\comp D
	\label{eq:090403c}
\end{eqnarray}
We calculate:
\begin{eqnarray*}
&&
	\meither A B\comp \msplit C D
\just={ (\ref{eq:msplit:def}) }
	\meither A B\comp (i_1\comp C + i_2\comp D)
\just={ bilinearity (\ref{eq:090403b}) }
	\meither A B\comp i_1\comp C + \meither A B\comp i_2\comp D
\just={ $+$-cancellation (\ref{eq:091206e})  }
	A\comp C + B\comp D
\end{eqnarray*}
Listing \ref{lst:matlab} converts this law into the corresponding \matlab\ algorithm 
for matrix multiplication.

As another example, let us show how standard block-wise
matrix-matrix multiplication (MMM),
\begin{eqnarray}
	\mblock R S T U
\comp
	\mblock A B C D
= 
	\mblock{R\comp A+S\comp C}{R\comp B+S\comp D}{T\comp A+U\comp C}{T\comp B+U\comp D}
	\label{eq:blockwise}
\end{eqnarray}
relies on \emph{divide-and-conquer} (\ref{eq:090403c}):
\begin{eqnarray*}
&&
	\meither{\msplit R T}{\msplit S U}\comp
	\meither{\msplit A C}{\msplit B D}
\just={ \emph{junc}-fusion (\ref{eq:meither:fusion}) }
	\meither{
		\meither{\msplit R T}{\msplit S U}\comp
		\msplit A C
	}{
		\meither{\msplit R T}{\msplit S U}\comp
		\msplit B D
	}
\just={ divide and conquer (\ref{eq:090403c}) twice }
	\meither{
		\msplit R T\comp A + \msplit S U\comp C
	~}{~
		\msplit R T\comp B + \msplit S U \comp D
	}
\just={ split-fusion (\ref{eq:meither:fusion}) four times }
%
%
	\meither{
		\msplit{R\comp A}{T\comp A} + \msplit{S\comp C}{U\comp C}
	~}{~
		\msplit{R\comp B}{T\comp B} + \msplit{S\comp D}{U\comp D}
	}
\just={ blocked addition (\ref{eq:abide:x}) twice }
	\meither{
		\msplit{R\comp A + S\comp C}{T\comp A + U\comp C}
	~}{~
		\msplit{R\comp B + S\comp D}{T\comp B + U\comp D}
	}
\just={ the same in block notation (\ref{eq:exc}) }
	\mblock{R\comp A+S\comp C}{R\comp B+S\comp D}{T\comp A+U\comp C}{T\comp B+U\comp D}
\end{eqnarray*}

\begin{lstlisting}[float=t,frame=tb,caption={Divide-and-conquer law (\ref{eq:090403c})
 converted to \matlab\ script for matrix-matrix multiplication. Blocks $A$, $B$ in
(\ref{eq:090403c}) are generated by partitioning argument matrix $X$ column-wise
and blocks $C$, $D$ are obtained in a similar way from $Y$. The algorithm
stops when both argument matrices degenerate into vectors ($k=1$). There
is no type checking, meaning that function \texttt{MMM} issues an error when the two
{\bf\texttt{size}} operations don't match --- the number of columns
(resp.\ lines) of $X$ (resp.\ $Y$) must be the same.},label=lst:matlab]
function R = MMM(X,Y)
    [k1, n] = size(Y);
    [m, k2] = size(X);
    if (k1 ~= k2) 
        error('Dimensions must agree');
    else
        k = k1;
        R = zeros(n, m);
        if k1 == 1
            R = X * Y;
        else
            k1 = round(k / 2);
            A = X(:,1:k1); B = X(:,k1+1:k);
            C = Y(1:k1,:); D = Y(k1+1:k,:);
            R = MMM(A,C) + MMM(B,D);
        end
    end
end
\end{lstlisting}

\section{Calculating Triple Nested Loops}
\label{sec:110107c}
By putting together the universal factorization of matrices in terms of the
\emph{junc} and \emph{split} combinators, one easily infers yet another such
property handling four blocks at a time:
\begin{eqnarray*}
X = \mblock {A_{11}} {A_{12}} {A_{21}} {A_{22}} & \equiv &
\left\{
	\begin{array}{rcl}
	\p1 \comp X \comp i_1 = {A_{11}}
	\\
	\p1 \comp X \comp i_2 = {A_{12}}
	\\
	\p2 \comp X \comp i_1 = {A_{21}}
	\\
	\p2 \comp X \comp i_2 = {A_{22}}
	\end{array}
\right.
\end{eqnarray*}
Alternatively, one may generalize (\ref{eq:meither:def},\ref{eq:msplit:def})
to blocked notation
\begin{eqnarray*}
\mblock {A_{11}} {A_{12}} {A_{21}} {A_{22}} = i_1 \comp A_{11} \comp \pi_1 + i_1 \comp A_{12} \comp \pi_2 + i_2 \comp A_{21} \comp \pi_1 + i_2 \comp A_{22} \comp \pi_2
\end{eqnarray*}
which rewrites to
\begin{eqnarray}
\mblock {A_{11}} {A_{12}} {A_{21}} {A_{22}} = \matrix {A_{11}} 0 0 0 + \matrix 0 {A_{12}} 0 0 + \matrix 0 0 {A_{21}} 0 + \matrix 0 0 0 {A_{22}}
	\label{eq:101221c}
\end{eqnarray}
once injections and projections are replaced by the biproduct solution of
Section \ref{sec:chasing}.

\paragraph{Iterated Biproducts} 
It should be noted that biproducts generalize to finitely many arguments,
leading to an $n$-ary generalization of the (binary) \emph{junc}/\emph{split}
combinators. The following notation is adopted in generalizing
(\ref{eq:meither:def},\ref{eq:msplit:def}):
\begin{eqnarray*}
	\left[\begin{array}{c|c|c} A_1 & \ldots & A_p \end{array}\right]
	\wider=
	\bigovert_{1\leq j\leq p} A_j
	\wider=
	\sum_{j=1}^{p} A_j \comp \pi_j
\\
	\left[\begin{array}{c} A_1 \\ \hline \vdots \\ \hline A_m \end{array}\right]
	\wider=
	\bigominus_{1\leq j\leq m} A_j
	\wider=
	\sum_{j=1}^{m} i_j \comp A_j
\end{eqnarray*}
Note that all laws given so far generalize accordingly to $n$-ary \emph{splits} and \emph{juncs}.
In particular, we have the following universal properties:
\begin{eqnarray}
	X= \bigovert_{1\leq j\leq p} A_j & \equiv & \bigwedge_{1\leq j\leq p} X  \comp i_j = A_j
	\label{eq:091207a}
\\
	X= \bigominus_{1\leq j\leq m} A_j & \equiv & \bigwedge_{1\leq j\leq m} \p j \comp X = A_j
	\label{eq:091207b}
\end{eqnarray}
The following rules expressing the block decomposition of a matrix $A$
\begin{eqnarray}
	   A=\left[\begin{array}{c|c|c} A_1 & \ldots & A_p \end{array}\right] & = & \bigovert_{1\leq j\leq p} A \comp i_j = \sum_{j=1}^{p} A \comp i_j \comp \pi_j
		\label{eq:iteeit:def}
\end{eqnarray}
\begin{eqnarray}
	   A=\left[\begin{array}{c} A_1 \\ \hline \vdots \\ \hline A_m \end{array}\right] & = & \bigominus_{1\leq j\leq m} \pi_j \comp A = \sum_{j=1}^{m} i_j \comp \pi_j \comp A
		\label{eq:itespl:def}
\end{eqnarray}
arise from the iterated definitions by letting $X=A$ in the universal properties
and substituting.

Further note that $m,p$ can be chosen as large as possible, the limit taking place when
blocks $A_i$ become atomic. In this limit situation,
a given matrix $\larrow n A m$ is defined in terms of its elements $A_{jk}$ as:
\begin{eqnarray}
	   A = \left[\begin{array}{c|c|c} a_{11} & \ldots & a_{1n}\\\hline \vdots & \ddots & \vdots \\ \hline  a_{m1} & \ldots & a_{mn} \end{array}\right] 
	& = &  \sum_{\substack{1\leq j\leq m \\1 \leq k \leq n }} i_j \comp \pi_j \comp A \comp i_k \comp \pi_k = \bigoplus_{\substack{1\leq j\leq m \\1 \leq k \leq n}} \pi_j \comp A \comp i_k 
		\label{eq:itebip:def}
\end{eqnarray}
where $\bigoplus_{\substack{1\leq j\leq m \\1 \leq k \leq n}}$ abbreviates $\bigominus_{1\leq j\leq m} \bigovert_{1 \leq k \leq n}$
--- equivalent to $\bigovert_{1 \leq k \leq n} \bigominus_{1\leq j\leq m}$ by the generalized exchange law (\ref{eq:exc}). 

Our final calculation shows how iterated biproducts ``explain'' the traditional
for-loop implementation of MMM. Interestingly enough, such iterative implementation
is shown to stem from generalized divide-and-conquer (\ref{eq:090403c}):
\begin{eqnarray*}
	C &=& A \comp B
\just={ (\ref{eq:itebip:def}), (\ref{eq:iteeit:def}) and (\ref{eq:itespl:def}) }
	(\bigominus_{1\leq j\leq m} \pi_j \comp A) \comp (\bigovert_{1 \leq k\leq n} B \comp i_k)
\just={ generalized split-fusion (\ref{eq:msplit:fusion}) }
	\bigominus_{1\leq j\leq m} (\pi_j \comp A \comp (\bigovert_{1 \leq k\leq n} B \comp i_k))
\just={ generalized either-fusion (\ref{eq:meither:fusion}) }
	\bigominus_{1\leq j\leq m} (\bigovert_{1 \leq k\leq n} \pi_j \comp A \comp B \comp i_k)
\just={ (\ref{eq:iteeit:def}), (\ref{eq:itespl:def}) and generalized (\ref{eq:msplit:fusion}) and (\ref{eq:meither:fusion}) }
	\bigominus_{1\leq j\leq m} (\bigovert_{1 \leq k\leq n} ((\bigovert_{1 \leq l\leq p} \pi_j \comp A \comp i_l) \comp (\bigominus_{1 \leq l\leq p} \pi_l \comp B \comp i_k)))	
%
%
\just={ generalized divide-and-conquer (\ref{eq:090403c}) }
	\bigominus_{1\leq j\leq m} (\bigovert_{1 \leq k\leq n} (\sum_{1 \leq l\leq p} \pi_j \comp A \comp i_l \comp \pi_l \comp B \comp i_k))	
\end{eqnarray*}

As we can see in the derivation path, the choices for the representation
of $A$ and $B$ impact on the derivation of the intended algorithm. Different
choices will alter the order of the triple loop obtained. 
Proceeding to the loop inference will involve the expansion of $C$ and the normalization of the formula
into sum-wise notation:  
\begin{eqnarray*}
&&
\bigoplus_{\substack{1 \leq k\leq m \\1 \leq j \leq n}} \pi_j \comp C \comp i_k = \bigominus_{1\leq j\leq m} (\bigovert_{1 \leq k\leq n} (\sum_{1 \leq l\leq p} \pi_j \comp A \comp i_l \comp \pi_l \comp B \comp i_k))	
\just\equiv{ (\ref{eq:itebip:def}), (\ref{eq:iteeit:def}) and (\ref{eq:itespl:def})}
\bigominus_{1\leq j\leq m} (\bigovert_{1 \leq k\leq n} \pi_j \comp C \comp i_k) = \bigominus_{1\leq j\leq m} (\bigovert_{1 \leq k\leq n} (\sum_{1 \leq l\leq p} \pi_j \comp A \comp i_l \comp \pi_l \comp B \comp i_k))
\end{eqnarray*}
At this point we rely on the universality of the \emph{junc} and  \emph{split} constructs
(\ref{eq:091207a},\ref{eq:091207b}) to obtain from above the post-condition of the algorithm:
\begin{eqnarray}
&&
\bigwedge_{1\leq j\leq m} (\bigwedge_{1 \leq k\leq n} (\pi_j \comp C \comp i_k = \sum_{1 \leq l\leq p} \pi_j \comp A \comp i_l \comp \pi_l \comp B \comp i_k))
\label{eq:110108a}
\end{eqnarray}

This predicate expresses an outer traversal indexed by $j$, an inner traversal
indexed by $k$ and what the expected result in each element of output
matrix $C$ is. Thus we reach three nested for-loops of two different
kinds: the two outer-loops (corresponding to indices $j,k$) provide for \emph{navigation}, while the inner loop
performs an \emph{accumulation} (thus the need for the initialization).

\begin{lstlisting}[frame=bt,float=h,caption={ \matlab\ encoding of naive
triple \texttt{for}-loop implementation of MMM, corresponding to traversing
the rows of $A$ through $j$ and the columns of $B$ via $k$. This is
a refinement of the calculated post-condition (\ref{eq:110108a}).},label=lst:NaiveMMM]
function C = NaiveMMM(A,B)

   [m, p1] = size(A);
   [p2, n] = size(B);
    
   if (p1 ~= p2) 
        error('Dimensions must agree');
   else
    for j = 1:m 
      for k = 1:n 
        C(j,k) = 0; 
        for l = 1:p1 
          C(j,k) = C(j,k) + A(j,l) * B(l,k); 
        end 
      end
    end
   end

end
\end{lstlisting}

Different matrix memory mapping schemes give rise to the interchange of the
$j,k$ and $l$ in the loops in Listing \ref{lst:NaiveMMM}.
(For a complete discussion of 
matrix partition possibilities see \cite{Goto:08}.)
This is due to corresponding choices in the derivation granted by the
generalized exchange law (\ref{eq:exc}), among others.

Other variants of blocked MMM (\ref{eq:blockwise}) such as e.g.\ Strassen's
or Winograd's \cite{Pao07} rely mainly on the additive structure of $\Matk$ and
thus don't pose new challenges.

\section{Developing biproduct algebra for applications}
\label{sec:110107d}
For a mathematical concept to be effective it should blend expressiveness 
with calculation power, while providing a generic setting wherefrom practically relevant
situations can be derived by instantiation.
It should also scale up, in the sense of exhibiting an algebra making it easy to 
``build new from old''.

We will see shortly that biproducts scale up in this manner.
So, instead of chasing new solutions to the biproduct equations 
and checking  which ``chapters'' of linear algebra \cite{MacLane99:Algebra}
they are able to constructively explain,
one may try and find rules which build new biproducts from existing ones so as to
fit into particular situations in linear algebra.

Think of Gaussian elimination, for instance,
whose main steps involve row-switch\-ing, row-multiplication and row-addition, and suppose
one defines the following transformation $t$ catering for the last two,
for a given $\alpha$:
\begin{eqnarray*}
	&& t : (\larrow n{}n) \times (\larrow{m}{}{n+n}) \rightarrow (\larrow{m}{}{n+n})
\\
	&& t(\alpha,\msplit A B) \wider= \msplit {A}{\alpha A + B}
\end{eqnarray*}
Thinking in terms of blocks $A$ and $B$ rather than rows is more general;
in this setting, arrow $\larrow n\alpha n$ means $\larrow n{id}n$ with all
$1$s replaced by $\alpha$s, and $\alpha A$ is $\alpha\comp A$.
Let us analyze transformation $t$ in this setting,
by using the blocked-matrix calculus in reverse order:
\begin{eqnarray*}
	t(\alpha,\msplit A B) &=& \msplit {A}{\alpha \comp A + B}
\just={ (\ref{eq:blockwise}) in reverse order }
	\mblock 1 0 \alpha 1 \comp \msplit A B
\just={ divide-and-conquer (\ref{eq:090403c}) }
	\msplit 1 \alpha \comp A + \msplit 0 1 \comp B
\end{eqnarray*}
It can be shown that the last expression, which has the same shape as
(\ref{eq:msplit:def}), is in fact the \emph{split} combinator generated by
another biproduct,
\begin{eqnarray*}
\begin{array}{lll}
	\p1' = \left[\begin{array}{cc} 1 & 0\end{array}\right]
&,&
	\p2' = \left[\begin{array}{cc}-\alpha & 1\end{array}\right]
\\[2ex]
	i_1' = \, \left[\begin{array}{cc} 1    \\ \alpha \end{array}\right]
&,&
	i_2' = \, \left[\begin{array}{cc}0 \\ 1\end{array}\right]
\end{array}
\end{eqnarray*}
parametric on $\alpha$.
In summary, this biproduct, which extends the one studied earlier on
(they coincide for $\alpha := 0$) provides a categorial interpretation 
of one of the steps of Gaussian elimination. 

Biproducts in $\Matk$ are unique up to isomorphism due to universality of
product and coproduct. 
\emph{Splitting} $\p1'$ and $\p2'$ with the standard projections (\ref{eq:101221a}) 
 is just another way to build \emph{elementary matrices} \cite{AM05} such as, for instance,
\begin{eqnarray*}
\msplit{\p1'}{\p2'}= i_1 \comp \p1' + i_2 \comp \p2' = 
\left[\begin{array}{cc} 1 & 0\\ -\alpha & 1\end{array}\right]
\end{eqnarray*}
which are central to Gaussian elimination. In essence,
this algorithm performs successive transformations of a given matrix
into isomorphic ones via elementary matrices 
that witness the isomorphisms.
%
%
Below we show that such elementary steps of Gaussian elimination scale up to
blocks via suitable biproduct constructions. The first one generalizes row switching to
block switching.

\begin{thm}[Swapping biproducts] Let
$ \xymatrix{
m \ar@<-.5ex>[r]_{i_1}  &   r \ar@<-.5ex>[l]_{\pi_1} \ar@<.5ex>[r]^{\pi_2}   & n \ar@<.5ex>[l]^{i_2}\\
}$ be a biproduct. Then swapping projections (resp.\ injections) with each other
yields another biproduct.

\textbf{Proof:} Obvious, as (\ref{eq:biprod:b}) swaps with (\ref{eq:biprod:a})
and (\ref{eq:biprod:c}) stays the same, since addition is commutative.
\end{thm}
For instance, swapping the standard biproduct yields another biproduct (superscript $s$ stands for \emph{swap}) :
\begin{eqnarray}
\begin{array}{ll}
i_1^s = \msplit 0 1 & \qquad i_2^s = \msplit 1 0
\\[3ex]
\p1^s = \meither 0 1 & \qquad \p2^s = \meither 1 0
\end{array}
	\label{eq:101124f}
\end{eqnarray}
Thus
\begin{eqnarray}
	\meither A B ^s & = & \meither B A
\\[2ex]
	\msplit A B ^s & = & \msplit B A
\end{eqnarray}
Swapped biproduct (\ref{eq:101124f}) generalizes row-swapping to block-swapping,
as the following example shows: the effect of swapping $A$ with
$B$ in matrix {\tiny 
\(
\begin{bmatrix} A \\ C \\ B
\end{bmatrix}
\) }
is obtained by representing it in \emph{swap mode}:
\begin{eqnarray*}
	\msplitwithSS A {\msplit C B^s}^s = \msplitwithSS{\msplit C B^s} A = \msplitwithSS{\msplit B C} A
=
\begin{bmatrix} B \\ \noalign{\smallskip} C \\ \noalign{\smallskip} A
\end{bmatrix}
\end{eqnarray*}

Next we want to show how to perform row-multiplication and addition at block-level.
There is a biproduct for this, also evolving from the standard:

\begin{thm}[Self cancellable biproducts]\label{th:111008b}
Replacing one of the $0$ components of projection $\p2$ (resp.\ $\p1$) of the standard biproduct (\ref{eq:101221a})
by an arbitrary (suitably typed) matrix $C$ and the corresponding $0$ component
of $i_1$ (resp.\ $i_2$) by
$-C$ yields a biproduct.
That is,
\begin{eqnarray}
\begin{array}{lll}
	\p1^C = \meither 1 0
&,&
	\p2^C = \meither {C} 1
\\[3ex]
	i_1^C = \msplit 1 {-C}
&,&
	i_2^C = \msplit 0 1
\end{array}
	\label{eq:101221b}
\end{eqnarray}
form a biproduct, parametric on $C$, where types are as in the diagram below:
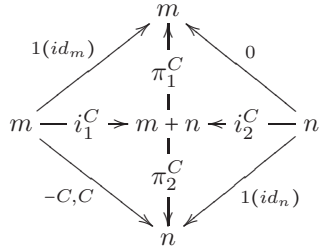
\begin{eqnarray*}
	\xymatrix@R+2ex@C+2ex{
&
	m  
&
\\
	m 
	\ar[dr]_{-C,C}
	\ar[ur]^{1 (id_m)}
	\ar[r]|(.45)*+{i_1^C} 
&
	m+n 
	\ar[u]|(.45)*+{\p1^C} 
	\ar[d]|(.45)*+{\p2^C}
&
	n 
	\ar[l]|(.45)*+{i_2^C} 
	\ar[dl]^{1 (id_n)} 
	\ar[ul]_{0} 
\\ 
&
	n  
	}
\end{eqnarray*}
\textbf{Proof:} See the appendix.
\end{thm}

Let us inspect the behaviour of the \emph{junc} (\ref{eq:meither:def})
and \emph{split} (\ref{eq:msplit:def}) combinators arising from this biproduct:
\begin{eqnarray}
	\meither A B ^C
&=&
	A \comp \pi_1^C + B \comp \pi_2^C
	\nonumber
\\ &=&
	A \comp \meither 1 0 + B \comp \meither {C} 1
\wider=
	\meither A 0 + \meither {B\comp C} B
	\nonumber
\\ &=&
	\meither {A+B\comp C} B
	\nonumber
\\
	\msplit  A B ^C
&=&
	i_1^C \comp A  + i_2^C \comp B
	\nonumber
\\ &=&
	\msplit 1 {-C} \comp A + \msplit 0 1 \comp B
\wider=
	\msplit A {-C \comp A} + \msplit 0 B
	\nonumber
\\ &=&
	\msplit A {(-C \comp A) + B}
\wider=
	\msplit A {B - C \comp A}
	\label{eq:110830a}
\end{eqnarray}
The universal property of \emph{split} will thus be:
\begin{eqnarray}
	X = \msplit  A B ^C
& \equiv &
	\p1 \comp X = A \wider\land
	\p2 \comp X + C \comp A = B
\end{eqnarray}
Note that $\msplitIL{A}{B}^{\mbox{\tiny $C$}}$ can be recognized as the block-version of an
operation common in linear algebra: replacing a row (cf.\ $B$) by subtracting
from it a multiple of another row (cf.\ $A$), as used in Gauss-Jordan elimination.
$\meitherIL{A}{B}^{\mbox{\tiny $C$}}$ does the same column-wise, adding rather than subtracting.

This enables the following
block-version of Gauss-Jordan elimination, where $X$ is supposed to be invertible
(always the case if in row-echelon form):
\begin{eqnarray}
\begin{array}{rcl}
&&
	\gje : (\larrow{k+m}{}{k+n}) \rightarrow (\larrow{k+m}{}{k+n})
\\
&&
\gje\ \mblock{
	X
}{
	 \MA{B}{1}{m}
}{
	\MA{A}{n}{1}
}{
	\MA{D}{n}{m}
}
	=
	 \left[\begin{array}{c|c} 
	X & \MA{B}{1}{m} \\
	\hline
	\MA{0}{n}{1}  & \gje(D - A \comp X^{-1} \comp  B)
	\end{array}\right] 
\\
&&
\gje\ \MA{X}{1}{1} = \MA{X}{1}{1}
\end{array}
	\label{eq:101221d}
\end{eqnarray}
$X^{-1}$ denotes the inverse of $X$, that is, $X\comp X^{-1}=id$ holds.
The rationale of the algorithm assumes that the swapping biproduct is first applied as
much as needed to transform the input matrix in the form
{\tiny \(
\mblock{
	X
}{
	 \MA{B}{1}{m}
}{
	\MA{A}{n}{1}
}{
	\MA{D}{n}{m}
}
\) }
where topmost-leftmost block $X$ is in row-echelon form.
(Listing \ref{lst:BGJE} shows an encoding of (\ref{eq:101221d}) into a \matlab\ script.)
Then the \emph{split} combinator (\ref{eq:110830a})
of the self-cancellable biproduct associated to $C=A \comp X^{-1}$ is used
to convert
{\tiny \(
\mblock{
	X
}{
	 \MA{B}{1}{m}
}{
	\MA{A}{n}{1}
}{
	\MA{D}{n}{m}
}
\) }
into a matrix in which cancellation ensures the $0$ block of
the right-hand side of  (\ref{eq:101221d}):
\begin{eqnarray*}
\msplitwithvskip {\meither X B}{\meither A D}^{(A \comp X^{-1})}
	&=&
\msplitwithSS {\meither X B}{\meither A D - (A \comp X^{-1}) \comp \meither X B}
\\
	&=&
\msplitwithSS {\meither X B}{\meither A D - \meither{A \comp X^{-1} \comp X}{A \comp X^{-1} \comp B}}
\\
	&=&
\msplitwithSS {\meither X B}{\meither{A - A}{D - A \comp X^{-1} \comp B}}
\\
	&=&
\mblock X B 0 {D-A \comp X^{-1} \comp B}
\end{eqnarray*}
The algorithm proceeds recursively applied to (smaller) block $D-A \comp X^{-1} \comp B$ until
$X$ is found alone, that is, the target type (i.e.\ number of rows) of $A$ and $D$ is 0.

\begin{lstlisting}[frame=bt,float=t,caption={ \matlab\ encoding of the algorithm
for blocked version of Gauss-Jordan elimination given by (\ref{eq:101221d}).
Auxiliary function \texttt{MPRef} calculates the size of the largest topmost-leftmost 
block of input matrix $M$ that is in row-echelon form.},label=lst:BGJE]
function R = GJE(M)
    [m,n] = size(M);
    k = MPRef(M);
    if k < n 
        X = M(1:k,1:k);
        B = M(1:k,k+1:n);
        A = M(k+1:m,1:k);
        D = M(k+1:m,k+1:n);
        R(1:k,1:k) = X;
        R(1:k,k+1:n) = B;
        R(k+1:m,1:k) = zeros(m-(k+1)+1,k);
        R(k+1:m,k+1:n) = GJE(D - A * inv(X) * B);
    else
        R = M;
    end
end
\end{lstlisting}

The classical version of the algorithm corresponds to making block
$\larrow k X k$
singular, $\larrow 1 x 1$, yielding
\begin{eqnarray}
\begin{array}{rcl}
&&
	ge : (\larrow{1+m}{}{1+n}) \rightarrow (\larrow{1+m}{}{1+n})
\\
&&
ge \mblock{
	x
}{
	 \MA{B}{1}{m}
}{
	\MA{A}{n}{1}
}{
	\MA{D}{n}{m}
}
	=
	 \left[\begin{array}{c|c} 
	x & \MA{B}{1}{m} \\
	\hline
	\MA{0}{n}{1}  & ge(D - \frac{A}{x} \comp B)
	\end{array}\right] 
\\
&&
ge\ \MA{x}{1}{1} = \MA{x}{1}{1}
\end{array}
	\label{eq:101123a}
\end{eqnarray}

The correction of the algorithm is discussed elsewhere \cite{Ma11} with
respect to the specification: \emph{transform the input matrix into one which
is in row-echelon (RE) form and keeps the same information}. In brief, (\ref{eq:101221d})
ensures RE-form since $X$ is in RE-form (by construction) and
$\gje(D - A \comp X^{-1} \comp B)$ inductively does so. The other requirement
is ensured by the universal properties underlying block-notation, granted
by the biproduct construction: \emph{splits} and \emph{juncs} are isomorphisms,
so they preserve the information of the blocks they put together. For instance,
denoting the hom-set of all matrices with $n$ columns and $m$ rows by $m^n$,
property (\ref{eq:091206c}) establishes isomorphism
\begin{eqnarray}
	m^n \times m^p & \wider\iso & m^{n+p}
\end{eqnarray}
--- cf.\ (34) on page 181 of \cite{MacLane99:Algebra}. So, all ``juncs''
of similarly typed matrices are isomorphic, meaning that they hold the same
information under different formats.

\paragraph{Scaling biproducts}
Finally, we address the operation of scaling a biproduct by some  factor
--- a device which will be required in
the calculational approach to vectorization of Section \ref{sec:101218b}.
The question is:
given biproduct
\(
\xymatrix{
	m \ar@<-.5ex>[r]_{i_1}
&	m+n \ar@<-.5ex>[l]_{\pi_1} \ar@<.5ex>[r]^{\pi_2}
&	n \ar@<.5ex>[l]^{i_2}
}
\),
can its dimensions be ``scaled up $k$ times''?

This will mean multiplying $m$ and $n$ (and $m+n$) by $k$.
The matrix operation which has this behaviour dimension-wise is the so-called
Kronecker product \cite{MN79}:
given $\larrow mAp$ and $\larrow nBq$, Kronecker product
$\larrow{m\times n}{A\otimes B}{p \times q}$ is the matrix
which replaces each element $a_{ij}$ of $A$ by block $a_{ij}B$.

In the categorial setting of $\Matk$, Kronecker product is a tensor product,
captured by a (bi)functor $\otimes : \Matk \times \Matk \rightarrow \Matk$.
On objects, $m \otimes n = m \times n$ (product of two dimensions); on morphisms,
$A \otimes B$ is the matrix product defined above.
Recall that a category is monoidal \cite{MacLane98:CategoriesWM,joyal1991geometry,joyal1996traced}
when it comes equipped with one such bifunctor
which is associative 
\begin{eqnarray}
	(A \otimes B) \otimes C & = & A \otimes (B \otimes C)
	\label{eq:tensor:assoc}
\end{eqnarray}
and has a left and a right unit. In the case of $\Matk$ the unit is $id_1$.
This means that we can rely on the following properties~\footnote{%
More can be said about $\Matk$ but for our purposes it is enough to stick
to its monoidal structure. Further properties can be found in \cite{dosen2005}.
For alternative definitions of the Kronecker product in terms of other matrix
products see section \ref{sec:110107f}.}
granting $\otimes$ as a bilinear bifunctor, for suitably typed $A$, $B$ and $C$:
\begin{eqnarray}
	(A \otimes B) \comp (C \otimes D) & = & (A \comp C) \otimes (B \comp D)
	\label{eq:tensor:functor}
\\
	id \otimes id & = & id
	\label{eq:101218c}
\\
	A \otimes ( B + C ) &=& (A \otimes B) + (A \otimes C)
	\label{eq:101218d}
\\
	( B + C ) \otimes A &=& (B \otimes A) + (C \otimes A)
	\label{eq:101218e}
\end{eqnarray}

\begin{thm}[Scaling biproducts]\label{th:111008c}
Let $ \xymatrix{ m \ar@<-.5ex>[r]_{i_1}  & m + n \ar@<-.5ex>[l]_{\pi_1} \ar@<.5ex>[r]^{\pi_2}   & n \ar@<.5ex>[l]^{i_2}\\ }$
be a biproduct. Then
\begin{eqnarray*}
\xymatrix@C=4em{
	m \times k
		\ar@<-.5ex>[r]_{i_1 \otimes {id_k}}
&	(m+n) \times k
		\ar@<-.5ex>[l]_{\pi_1 \otimes {id_k}}
		\ar@<.5ex>[r]^{\pi_2\otimes {id_k}}
&	n\times k
		\ar@<.5ex>[l]^{i_2\otimes {id_k}}
}
\end{eqnarray*}
is a biproduct.

\textbf{Proof:} See the appendix.
\end{thm}

\noindent
This result has a number of nice consequences,
namely two simplification rules
\begin{eqnarray}
	\p j \otimes id & = & \p j ~~~~ (j=1,2)
	\label{eq:101218f}
\\
	i_j \otimes id & = & i_j ~~~~ (j=1,2)
	\label{eq:111008d}
\end{eqnarray}
which lead to the two Kronecker-product fusion laws,
\begin{eqnarray}
	\meither A B \otimes C &=& \meither{A \otimes C}{B \otimes C}
	\label{eq:101218g}
\\
	\msplit A B \otimes C &=& \msplit{A \otimes C}{B \otimes C}
	\label{eq:111007d}
\end{eqnarray}
which in turn provide for blocked Kronecker product operation.
The simplification rules are better understood with types made explicit, for instance
\begin{eqnarray*}
	(\larrow{m+n}{\p1}n) \otimes (\larrow k {id} k) & = & \larrow{k\times m+k\times n}{\p1}{k\times n}
\end{eqnarray*}
thus exhibiting the type polymorphism of biproduct injections and projections.
The calculation of fusion law (\ref{eq:101218g}) is given in the
appendix and that of (\ref{eq:111007d}) is similar.

Finally, we define another \( \Matk \) bifunctor --- \emph{direct sum},
\begin{eqnarray}
	A \oplus B & = & \meither {i_1 \comp A} {i_2 \comp B}
	\label{eq:101222a}
\end{eqnarray}
of type
\begin{eqnarray*}
\xymatrix{
	n
		\ar[d]_A
&
	m
		\ar[d]_B
&
	n+m
		\ar[d]^{A \oplus B}
\\
	k
&
	j
&
	k+j
}
\end{eqnarray*}
which is such that
\begin{eqnarray}
	id_2 \otimes A &=& A \oplus A
	\label{eq:101221e}
\end{eqnarray}
holds.
From (\ref{eq:101222a}) we see that each biproduct generates its own direct sum.
This offers a number of standard properties which can be expressed
using coproduct (dually product) combinators. Thus absorption-law
\begin{eqnarray}
	\meither A B \comp (C \oplus D) &=& \meither {A \comp C}{B \comp D}
	\label{eq:101221f}
\end{eqnarray}
and the injections' natural properties which follow:
\begin{eqnarray}
	(A \oplus B) \comp i_1 &=& i_1 \comp A
\\
	(A \oplus B) \comp i_2 &=& i_2 \comp B	\label{eq:injnatprop}
\end{eqnarray}
The same properties can be expressed by reversing the arrows, that is, in terms
of projections and products. Checking them all from (\ref{eq:101222a}) and
the universal property of \emph{junc} (dually: \emph{split}) is routine work.

\section{Vectorization: ``from product to exponentiation''}
\label{sec:101218b}
Vectorization (or linearization) is the operation (linear transformation)
which converts a matrix into a (column) vector~\footnote{
``Vectorization'' is an ambiguous term, for it also means using SIMD vector
instructions \cite{Pueschel:11} and not storing matrices as vectors. We adhere to it because
of its widespread use
in the bibliography, see eg.\ \cite{MN79,AM05,Sc10}.}. Given matrix $A$ below, we
can transform it into vector $v$ as
shown, which corresponds to parsing $A$ in column-major order :\\
\begin{minipage}{0.5\linewidth}
\centering
	\[
	A=\begin{bmatrix}a_{11} & a_{12} & a_{13}\\a_{21} & a_{22} & a_{23}\end{bmatrix}
	\]
\end{minipage}
\begin{minipage}{0.5\linewidth}
\centering
	\[
	\vect A = \begin{bmatrix}a_{11} \\ a_{21} \\ a_{12}\\a_{22} \\ a_{13} \\ a_{23}\end{bmatrix}
	\]
\end{minipage}

The linearization of an arbitrary matrix into a vector is a data refinement step.
This means finding suitable abstraction/representation relations \cite{Ol08a}
between the two formats and reasoning about them, including the refinement
of all matrix operations into vector form. In this section we show that
such an abstraction/representation pair is captured by isomorphisms
implicit in a universal construct, and use these in calculating
the implementation of two matrix combinators --- composition and transpose.

\subsection{Column-major Vectorization}
In the example given above, matrix $A$ is of type $\larrow 3{}2$ and
$\vect A$ is of type $\larrow 1 {} 6$. So, we can write the type of operator $\vect$ as
follows:
\[
   \vect :: (2 \leftarrow 3 \times 1) \rightarrow  (3 \times 2 \leftarrow  1)
\]
Writing $3\times 1$ (resp.\ $3\times 2$) instead of $3$ (resp.\ $6$)
is suggestive of the polymorphism of this operator,
\[
	\vectk k {} :: (n \leftarrow k \times m) \rightarrow  (k \times n \leftarrow  m)
\]
where a factor $k$ is shunted between the input and the output types.

Thus vectorization is akin to exponentiation, that is, \emph{currying} \cite{Ha03}
in functional languages. While currying ``thins'' the input of a given binary
function $f :: c \from a \times b$ by converting it into its unary (higher-order) counterpart
$\curry f :: (c \from b) \from a$, so does
vectorization by thinning a given matrix $\larrow{k\times m}An$ into $\larrow {m}{\vect
A}{k\times n}$.

We will refer to $k$ as the ``thinning factor'' of the vectorization.
This factor is $k=3$ in the illustration above. For
$m=1$, $\vect A$ becomes a column vector: the standard situation considered
in the literature \cite{MN79,AM05}.

As we shall see briefly, operator $\vectk k$ is a bijection, in fact one of the
witnesses of the isomorphism that underlies the empirical
observation that vectorization and devectorization preserve matrix contents,
changing matrix shape only. The other witness is its converse $\unvectk k$:
\[
\xymatrix{
n \leftarrow k \times m \ar@/^1pc/[rr]^{\vectk k {}}  & \cong & k \times n \leftarrow  m \ar@/^1pc/[ll]^{\unvectk k {}}
}
\]

As we did for other matrix combinators, we shall capture such intuition formally
in the form of the universal property which wraps up the isomorphism above,
this time finding inspiration in \cite{DP03}:
\begin{eqnarray}
	X = \vectk k A \wider\equiv A = {\epsilon_k} \comp  (id_k \otimes X)
& ~ &
\xymatrix{
	k \times n
 & 
	k \times (k \times n)
		\ar[r]^-{{\epsilon_k}}
 & 
	n
 &
\\
	m
		\ar[u]^{X}
&
	k \times m
		\ar[u]^{id_k \otimes X}
		\ar[ru]_{A}
}
	\label{eq:101124a}
\end{eqnarray}

Following the standard recipe, (\ref{eq:101124a}) grants $\vect{}$ and its
converse $\unvect{}$ as bijective transformations.
Among the usual corollaries of (\ref{eq:101124a}) we record the following, which
will be used shortly: the cancellation-law,
\begin{eqnarray}
    A = \epsilon_k \comp (id_k \otimes \vectk k A)
	\label{eq:adj:vec:canc????}
	\label{eq:101219b}
\end{eqnarray}
obtained for $X := \vectk k A$, and a closed formula for devectorization,
\begin{eqnarray}
	\unvect X &= & \epsilon \comp (id \otimes X)
	\label{eq:101227a}
\end{eqnarray}
obtained from (\ref{eq:101124a}) knowing that $X = \vect A$ is the same as
$\unvect X= A$.

For $k=1$ it is easy to see that vectorization degenerates into identity:
$\vectk 1 A =A$ and $\epsilon_1=id$. We start by putting our index-free, biproduct matrix
algebra at work in the calculation of $\epsilon_k$ for $k=2$.

\paragraph{Blocked vectorization}
For $k=2$, the smallest possible case happens
for $m=n=1$, where one expects $\vectk 2{\begin{bmatrix}x\ y\end{bmatrix}}$ to be
{\tiny $\begin{bmatrix}x\\y\end{bmatrix}$}, for $x$ and $y$ elementary data.
We proceed to the generalization of this most simple situation by replacing $x$ and $y$ with
blocks
$\larrow m A n$ and $\larrow m B n$, respectively,
and reasoning:
\begin{eqnarray*}
&&
	\vectk 2 {\meither A B} \wider= \msplit A B
\just\equiv{ (\ref{eq:101124a}) }
	\meither A B \wider= {\epsilon_2} \comp (id_2 \otimes \msplit A B)
\just\equiv{ (\ref{eq:101221e}) ; unjunc ${\epsilon_2}$ into ${\epsilon_2}_1$ and ${\epsilon_2}_2$ }
	\meither A B \wider= \meither{{\epsilon_2}_1}{{\epsilon_2}_2} \comp (\msplit A B \oplus \msplit A B) 
\just\equiv{ $\oplus$-absorption (\ref{eq:101221f}) }
	\meither A B \wider=
	\meither{{\epsilon_2}_1 \comp \msplit A B}{{\epsilon_2}_2 \comp \msplit A B}
\just\equiv{ (\ref{eq:101221g}) ; (\ref{eq:101221h}) }
	{\epsilon_2}_1 = \p1 \land {\epsilon_2}_2 = \p2
\just\equiv{ junc  ${\epsilon_2}_1$ and ${\epsilon_2}_2$ back into  ${\epsilon_2}$ }
	{\epsilon_2} = \meither{\p1}{\p2}
\end{eqnarray*}
We have obtained, with types
\begin{eqnarray}
	\larrow{2n+2n}{\epsilon_2}n \wider= \meither{\p1}{\p2}
	\label{eq:101221i}
\end{eqnarray}
expressing $\epsilon_k$ (for $k=2$) in terms of the standard biproduct projections.
Thus $\vectk 2 \epsilon = \msplitIL{\p1}{\p2} = id_2$,
a particular case of \emph{reflection law},
\begin{eqnarray}
    \vectk k {\epsilon_k} &=&id_{k\times n}
	\label{eq:101227e}
\end{eqnarray}
easy to obtain in general from (\ref{eq:101124a}) by letting $X := id_{k\times n}$
and simplifying. This can be rephrased into
\begin{eqnarray}
{\epsilon} &=& \unvect{id}
	\label{eq:adj:vec:canc}
\end{eqnarray}
providing a generic way of defining the mediating matrix $\epsilon$ in (\ref{eq:101124a}).

As an exercise, we suggest the reader checks the following instance of
cancellation law (\ref{eq:adj:vec:canc}), for $k=m=n=2$:
\[
\begin{bmatrix}a_{11} & a_{12} \\a_{21} & a_{22}\end{bmatrix} = 
	\begin{bmatrix}1 & 0 & 0 & 0 & 0 & 0 & 1 & 0 \\0 & 1 & 0 & 0 & 0 & 0 & 0 & 1 \end{bmatrix} \comp (id_2 \otimes \begin{bmatrix}a_{11} \\ a_{21} \\a_{12} \\ a_{22}\end{bmatrix}  )
\]
It can be observed that
	$\epsilon = \begin{mysmatrix}1 & 0 & 0 & 0 & 0 & 0 & 1 & 0 \\0 & 1 & 0 & 0 & 0 & 0 & 0 & 1 \end{mysmatrix}
	= \begin{bmatrix}1 & 0 & 0 & 1 \end{bmatrix} \otimes id_2$.
This illustrates equality
\begin{eqnarray}
	\epsilon \otimes id &=& \epsilon
	\label{eq:101218fh}
\end{eqnarray}
easy to draw from previous results
\footnote{See the appendix. Equality (\ref{eq:101218fh})  provides an explanation for
the index-wise construction of $\epsilon$ given in \cite{DP03}.}.
Again rendering types explicit helps in checking what is going on:
\begin{eqnarray*}
	(\rarrow{k^2\times n}{\epsilon_k}{n}) \otimes (\rarrow j {id} j)
	& = &
	\rarrow{k^2\times (n\times j)}{\epsilon_k}{n\times j}
\end{eqnarray*}
Doing a similar exercise for $k=3$,
\(
	\vectk 3 \begin{bmatrix}A & B & C\end{bmatrix} =
	\begin{mysmatrix}A \\ B \\ C\end{mysmatrix}
\)
--- that is,
\begin{eqnarray*}
	\vectk 3 \meither{\meither A B} C = \msplitwithSS{\msplit A B} C
\end{eqnarray*}
---
one would obtain for
$\rarrow{3^2\times n}{\epsilon_3}{n}$ matrix
$\meitherIL{\meither{\p1\comp\p1}{\p2\comp\p1}}{\p2}$,
with types as in diagram:
\begin{eqnarray*}
\xymatrix{
	n
&
	n+n
		\ar[l]_{\p1}
		\ar[d]_{\p2}
&
	(n+n) + n
		\ar[l]_{\p1}
		\ar[r]^-{\p2}
&
	n
\\
&
	n
}
\end{eqnarray*}

Recalling absorption law  (\ref{eq:101221f}), (\ref{eq:101221e}) and $\epsilon_2$ (\ref{eq:101221i}),
we observe that $\epsilon_3$ rewrites to
\( 
	\meitherIL{\meither{\p1}{\p2}\comp(\p1\oplus\p1)}{\p2}
\),
itself the same as
\( 
	\meitherIL{\epsilon_2\comp(id_2\otimes \p1)}{\p2}
\),
providing a hint of the general case:
\begin{eqnarray}
	\epsilon_{k+1} &=& \meither{
		 {\epsilon_k \comp (id_k \otimes \p1)}}
		{\p2}
	\label{eq:101221j}
\\
	\epsilon_1 &=&id
	\label{eq:101224b}
\end{eqnarray}

Let us typecheck (\ref{eq:101221j}), assuming completely independent types as starting point:
\begin{eqnarray*}
	&& \larrow{(k+1)^2\times j}{\epsilon_{k+1}}j
\\
	&& \larrow{k^2\times i}{\epsilon_k}i
\\
	&& \larrow{k}{id_k}{k}
\\
	&& \larrow{n+m}{\p1}n
\\
	&& \larrow{a+b}{\p2}b
\end{eqnarray*}
Type equations $a=n$ and $b=m$ follow from $\p1$ and $\p2$ belonging to the same biproduct.
Term $\epsilon_k \comp (id_k \otimes \p1)$ forces type equation (``unification'')
\(
	k^2\times i = k\times n
\), that is, $n=k\times i$.
Term $\meitherIL{\epsilon_k \comp (id_k \otimes \p1)} {\p2}$ entails $i=m$.
Finally, the whole equality forces
\begin{eqnarray*}
	j&=&m
\\
	(k+1)^2\times j &=&k\times(k\times i+m) + m + k\times i
\end{eqnarray*}
whereby, unfolding and substituting,
$k^2\times m+2k\times m+m = k^2\times i+k\times m+i+k\times i$ yields $i=m$.
Thus the most general types of the components of (\ref{eq:101221j}) are:
\begin{eqnarray*}
	&& \larrow{(k+1)^2\times m}{\epsilon_{k+1}}m
\\
	&& \larrow{k^2\times m}{\epsilon_k}m
\\
	&& \larrow{k}{id_k}{k}
\\
	&& \larrow{k\times m+m}{\p1}{k\times m}
\\
	&& \larrow{k\times m+m}{\p2}m
\end{eqnarray*}
as displayed in the following diagram (dropping $\times$ symbols for better layout):
\begin{eqnarray*}
\xymatrix@C=2ex@R=1ex{
	k(km+m)
		\ar[rr]^-{i_1}
		\ar[rd]_{id_k\otimes\p1}
&
&
	(k+1)^2m=k(km+m)+(km+m)
		\ar[dd]|(.45)*^{\epsilon_{k+1}}
&
&
	km+m
		\ar[ll]_-{i_2}
		\ar[lldd]^{\p2}
\\
&
	k(km)
		\ar[rd]_{\epsilon_{k}}
\\
&
&
	m
}
\end{eqnarray*}
From (\ref{eq:101227a}) and looking at the diagram above we find an
even simpler way of writing
(\ref{eq:101221j}):
\begin{eqnarray*}
	\epsilon_{k+1} &=& \meither{\unvectk k {\p1}}{\p2}
\end{eqnarray*}

Summing up, the index-free definition of the counit $\epsilon_k$
of a vectorization for any thinning factor $k$ is made possible by use of
the biproduct construction, by induction on $k$. The
corresponding encoding in \matlab\ can be found in Listing \ref{lst:101224a}.

\begin{lstlisting}[frame=bt,float=t,caption={ \matlab\ encoding of
	$\epsilon_k$ (\ref{eq:101221j},\ref{eq:101224b}). Auxiliary function \texttt{jay} (cf.\ letter J)
	implements biproduct components $\p2$ and $i_2$ in the same way
	\texttt{eye} (cf.\ letter I) implements $\p1$ and $i_1$.
Both \texttt{eye} and \texttt{kron} are primitive operations in \matlab\ providing the identity
matrix and the Kronecker product operation.
},label=lst:101224a]
function E = epsilon(k,m)
    if k==1
       E = eye(m)
    else
       n=k-1;
       p1 = eye(n*m,k*m);
       p2 = jay(m,k*m);
       E = [ (epsilon(n,m) * kron(eye(n),p1)) p2 ]
    end
end

function J = jay(r,c)
         if r>=c 
            J = [ zeros(r-c,c) ; eye(c) ];
         else
            J = [ zeros(r,c-r) eye(r) ];
         end
end
\end{lstlisting}

\subsection{Devectorization}
There is another way of characterizing column-major vectorization, and this
is by reversing the arrows of (\ref{eq:101124a}) and expressing the universal
property of $\unvect$, rather than that of $\vect$,
\begin{eqnarray}
	X = \unvectk k A \wider\equiv A =  (id_k \otimes X) \comp \eta_k
	\label{eq:adj:unvec:univ}
\end{eqnarray}
cf.\ diagram
\hfill
\(
\xymatrix@C=3ex{
	k \times n
		\ar[d]_{X}
&
	k \times (k \times n)
		\ar[d]_{id_k \otimes X }
 & 
	n
		\ar[l]_-{\eta_k}
		\ar[ld]^{A}
\\
	m
&
	k \times m
\\
}
\)
\hfill\rule[-4.7em]{4em}{0pt}\\
where (dropping subscripts) $\eta = \conv\epsilon$ \cite{DP03}. From this we infer
not only the cancellation-law of devectorization,
\begin{eqnarray}
    A =  (id \otimes \unvect A) \comp \eta
	\label{eq:adj:unvec:canc}
\end{eqnarray}
but also a closed formula for vectorization,
\begin{eqnarray}
	\vect X & = & (id \otimes X) \comp \eta
	\label{eq:101224c}
\end{eqnarray}
since $X = \unvect A$ is the same as $\vect X = A$.
Thus
\begin{eqnarray}
\eta_k &=& \vect_k \, id_{k \times m}
\label{eq:111008e}
\end{eqnarray}
holds.
Reversing the arrows also entails the following converse-duality,
\begin{eqnarray}
	\conv{(\vect A)} &=& \unvect{(\conv A)}
        \label{eq:vectConv}
\end{eqnarray}
easy to draw from (\ref{eq:101227a}) and (\ref{eq:101224c}).

\subsection{Self-adjunction}
Summing up,
we are in presence of an adjunction between functor ${\fun F} X = id_k \otimes X$
and itself --- a \emph{self-adjunction} \cite{DP03} --- inducing a monoidal
closed structure in the category.
The root for this is again the biproduct,
entailing the same functor $\oplus$ (\ref{eq:101222a}) for both coproduct and
product. It is known that the latter is the right adjoint of the diagonal
functor $\Delta(n)=(n,n)$, which in turn is the right adjoint of the former.
Using adjunction's notation, $\oplus \dashv \Delta$ and
$\Delta \dashv \oplus$ hold. By adjunction composition \cite{MacLane98:CategoriesWM}
one obtains
$(\oplus \comp \Delta) \dashv (\oplus \comp \Delta)$, whereby
--- because $\oplus \comp \Delta = (id_2 \otimes\relax)$ (\ref{eq:101221e}) ---
the self-adjunction $(id_2 \otimes) \dashv (id_2 \otimes)$ holds.

\section{Unfolding vectorization algebra}
\label{sec:110107e}
This section will show how vectorization theory,
as given in eg.\ \cite[Chap.~10]{AM05}, follows from universal properties
(\ref{eq:101124a},\ref{eq:adj:unvec:canc}) by index-free calculation.
This is an advance over the traditional, index-wise matrix representations
and proofs \cite{MN79,DP03,AM05} where notation is often quite loose,
full of dot-dot-dots.
We will also stress on the role of matrix types in the reasoning.

\paragraph{Vectorization is linear}
To warm up let us see a rather direct result, the linearity of $\vect$:
\begin{eqnarray}
	\vect(A+B) &=& \vect A + \vect B	\label{eq:vecOfSum}
\end{eqnarray}
Its derivation,
which is a standard exercise in algebra-of-programming calculation style,
can be found in the appendix.

\paragraph{Roth's relationship}
On the other side of the spectrum we find
the following relationship of the $\vect$ operator and Kronecker product
\begin{eqnarray}
	\vect{(A \comp B \comp C)} & = & (\conv C \otimes A) \comp \vect B
	\label{eq:101124b}
\end{eqnarray}
which Abadir and Magnus \cite{AM05}
attribute to Roth \cite{Ro34} and regard as the fundamental result of the whole theory.

In \cite{AM05}, (\ref{eq:101124b}) is said to hold ``whenever the product $ABC$ is defined''.
Our typed approach goes further in  enabling us to find 
the \emph{most general} type which accommodates the equality. The exercise is worthwhile detailing
in so far as it spells out two different instances of polymorphic $\vect{}$,
with different thinning-factors. We speed up the inference by starting from types
already equated by the matrix compositions and by the equality as a whole:
\begin{eqnarray*}
\xymatrix@C=8ex@R=1pt{
	j
&
	n
		\ar[l]_{A}
&
	k
		\ar[l]_{B}
&
	m
		\ar[l]_{C}
\\
	m \times j
&
	k \times n
		\ar[l]_{\conv C \otimes A}
&
	x
		\ar[l]_{\vect B}
\\
	m \times j
&&
	x
		\ar[ll]^{\vect{(A\comp B \comp C)}}
}
\end{eqnarray*}
The type relationship between $B$ and $\vect B$ entails $k=k\times x$, and therefore $x=1$.
Thus the \emph{principal type} of (\ref{eq:101124b}) is:
\begin{eqnarray*}
\xymatrix@C=8ex@R=1pt{
	j
&
	n
		\ar[l]_{A}
&
	k
		\ar[l]_{B}
&
	m
		\ar[l]_{C}
\\
	m \times j
&
	k \times n
		\ar[l]_{\conv C \otimes A}
&
	1
		\ar[l]_{\vectk k B}
		\ar@/^1pc/[ll]^{\vectk m{(A\comp B \comp C)}}
}
\end{eqnarray*}

We will show briefly that (\ref{eq:101124b}) is the merge of two other facts which express
the vectorization of the product of two matrices $B$ and $C$ in two alternative ways,
\begin{eqnarray}
	\vectk k{(B \comp C)} &=& (id_k \otimes B) \comp \vectk k{C}
	\label{eq:101124c}
\\
	\vectk m (C\comp B) &=& (\conv B \otimes id_n) \comp \vectk k C
	\label{eq:101124d}
\end{eqnarray}
whose types schemes are given by diagrams
\begin{eqnarray}
\xymatrix@C=8ex@R=1pt{
	j
&
	n
		\ar[l]_{B}
&
	k\times m
		\ar[l]_{C}
\\
	k \times j
&
	k \times n
		\ar[l]_{id_k \otimes B}
&
	m
		\ar[l]_{\vectk k C}
		\ar@/^1pc/[ll]^{\vectk k{(B\comp C)}}
}
	\label{eq:110103a}
\end{eqnarray}
and
\begin{eqnarray*}
\xymatrix@C=8ex@R=1pt{
&
	n
&
	k
		\ar[l]_{C}
&
	m
		\ar[l]_{B}
\\
	m \times n
&
	k \times n
		\ar[l]_{\conv B \otimes id_n}
&
	1
		\ar[l]_{\vectk k C}
		\ar@/^1pc/[ll]^{\vectk m{(C\comp B)}}
}
\end{eqnarray*}
respectively.
The derivation of (\ref{eq:101124c}) follows by instantiation of cancellation
law (\ref{eq:adj:unvec:canc}), for $A := \vect{(B \comp C)}$, knowing that
$\unvect{(\vect X)} = X$ --- see the appendix.
The calculation of (\ref{eq:101124d}), also in the appendix, proceeds in the same manner.
Thanks to these two results, calculating (\ref{eq:101124b}) is routine work:
\begin{eqnarray*}
&&
	(\conv C \otimes A) \comp \vect B
\just={ identity ; bifunctor $\otimes$}
	(\conv C \otimes id) \comp (id \otimes A) \comp \vect B
\just={ (\ref{eq:101124c}) }
	(\conv C \otimes id) \comp \vect (A \comp B)
\just={ (\ref{eq:101124d}) }
	\vect (A \comp B \comp C)
\end{eqnarray*}

Roth's relationship (\ref{eq:101124b}) is proved in different ways in the
literature. In \cite{MN79}, for instance, it turns up in the proof of a result
about the \emph{commutation matrix} which will be addressed in the following
section. In \cite{AM05} it is calculated by expressing matrix $B$ as a
summation of vector compositions and relying on the linearity of $\vect\relax$,
using an auxiliary result about Kronecker product of vectors.  In a similar
approach, a proof for the particular case of boolean matrices is presented in
\cite{Sc10} using relational product. 

\paragraph{Vectorization as (blocked) transposition}
Finally, we state a result which relates vectorization with transposition
--- compare with (\ref{eq:110103e}):
\begin{eqnarray}
 \vectk {k+k'}{\meither A B} &=& \msplit{\vectk k A}{\vectk{k'} B}
	\label{eq:110103d}
\end{eqnarray}
Type inference reveals that the most generic types which accommodate this
result are $\larrow {k\times m}An$ and $\larrow {k'\times m}Bn$.
The proof can be found in the appendix.

When does, then, vectorization (\ref{eq:110103d})  \emph{coincide} with transposition (\ref{eq:110103e})?
We reason:
\begin{eqnarray*}
&&
	\vectk {k+k'}{\meither A B} = \conv{\meither A B}
\just\equiv{ (\ref{eq:110103e}) ; (\ref{eq:110103d}) ; (\ref{eq:110103f}) }
	\vectk{k}{A} = \conv A \wider\land \vectk{k'}{B} = \conv B
\end{eqnarray*}
The two clauses correspond to the induction hypothesis in a structurally
inductive argument, breaking down thinning factors until base case $k=1$
is reached. Since $\vectk 1 X = X$, we conclude that vectorization \emph{is}
transposition wherever $A$ and $B$ can be broken in ``rows'' of symmetric blocks,
that is, blocks $X$ such that $X=\conv X$. In the particular
case of \meitherIL{A}{B} being a row vector (type $n=1$), this always happens,
the symmetric blocks being individual cells of type $\larrow 1 \relax 1$.
Thus
\begin{eqnarray}
	\vectk{m} A &=&	\conv A
	\label{eq:vectorTransp}
\end{eqnarray}
holds, for $\larrow m A 1$ a row vector.

\section{Calculating vectorized operations}
\label{sec:101229d}
We close the paper by showing how typed linear algebra helps in calculating
matrix operations in vectorial form. We only address the two basic combinators
\emph{transpose} and \emph{composition}, leaving aside the sophistication
required by the parallel implementation of such combinators.
(See eg.\ \cite{Johnson:90,Pueschel:05,TSoPMC,Franchetti:09,Pueschel:11}
concerning the amazing evolution of the subject in recent years.)

To begin with, let us show that transposition can be expressed solely in terms of the $\vect\relax$ and
$\unvect\relax$ combinators.
The argument is a typical example of reasoning with arrows in
a categorial framework.
Let $\larrow mAn$ be an arbitrary matrix. We start by building
$\longlarrow{n\times m}{B=\unvectk n A}1$, then
$\longlarrow1{C=\vectk{n\times m} B}{n\times m}$ and, finally
$\longlarrow n{\unvectk{n} C}{m}$.
So, the outcome $\unvectk{n}{(\vectk{n\times m} {(\unvectk n A)})}$
has the same type as $\conv A$.
Checking that they are actually the same arrow
is easy, once put in another way:
\begin{eqnarray}
	\vectk n{(\conv A)} &=& \vectk{n\times m} (\unvectk n A)
	\label{eq:101227c}
\end{eqnarray}
We calculate: 
\begin{eqnarray*}
&&
	\vectk{n}{(\conv A)} 
\just={ (\ref{eq:vectConv})}
%
%
	\conv{(\unvectk{n}{A})}
%
\just={ (\ref{eq:vectorTransp}) since $\unvectk{n}{A}$ is of type $\larrow {m\times n} \relax 1$}
	\vectk{m \times n}{(\unvectk{n}{A})}
\end{eqnarray*}
Next, we show how (\ref{eq:101227c}) helps in calculating a particular,
generic matrix --- the \emph{commutation matrix} --- usefull to implement
transposition of matrices encoded as vectors using
matrix-vector products.

\subsection{Implementing transposition in vectorial form} 
Magnus and Neudecker \cite{MN79} present the \emph{commutation matrix} $\larrow{m \times n}{K_{nm}}{n \times m}$
as the unique solution $K_{nm}$ to equation 
\begin{eqnarray}
	\vectk n{(\conv A)} = K_{nm} \comp \vectk m A
&
	\mbox{~ cf.\ diagram ~}
&
\xymatrix@C=2.3ex{
	n\times m
&
	m \times n
		\ar[l]_{K_{nm}}
&
	n
\\
&
	1
		\ar[ul]^{\vectk n{\conv A}}
		\ar[u]_{\vectk m A}
&
	m
		\ar[u]_{A}
}
	\label{eq:101227d}
\end{eqnarray}
the practical impact of which is obvious: knowing how to
build (generic) $K_{nm}$ enables one to transpose matrix $A$ by
composing $K_{nm}$ with $A$ vectorized, the outcome being delivered
as a vector too. Implemented in this way, transposition can
take advantage of the \emph{divide-and-conquer} nature of matrix multiplication
and therefore be efficiently performed on parallel machines.

The uniqueness of $K_{nm}$ is captured by the ``universal'' property,
\begin{eqnarray}
	X = K_{nm} & \equiv & \vect(\conv A) = X \comp \vect A
	\label{eq:101124e}
\end{eqnarray}
of which (\ref{eq:101227d}) is the cancellation corollary.
However, (\ref{eq:101124e}) defines $K_{nm}$ implicitly, not its explicit
form.  In the literature, this matrix (also referred to as
the \emph{stride permutation} matrix \cite{Johnson:90,Voronenko:08})
is usually given using indexed notation. For instance, Magnus
and Neudecker \cite{MN79} define it as a double summation
\begin{eqnarray}
	K_{nm}&=& \sum_{i=1}^n \sum_{j=1}^m(H_{ij}\otimes\conv{H}_{ij})
	\label{eq:101229c}
\end{eqnarray}
where each component $H_{ij}$ is a $(n,m)$ matrix with a $1$ in its $ij$th
position and zeros elsewhere.

Below we give a simple calculation of its generic formula, arising from
putting  (\ref{eq:101227d}) and (\ref{eq:101227c}) together:
\begin{eqnarray*}
	K_{nm} \comp \vectk m A
	&=&
	\vectk{n\times m} (\unvectk n A)
\end{eqnarray*}
Knowing the reflection law $\vectk m {\epsilon_m} = id$ (\ref{eq:101227e}) and substituting
we obtain a closed formula for the commutation matrix:
\begin{eqnarray}
	K_{nm} &=& \vectk{n\times m}{(\unvectk n{\epsilon_m})}
	\label{eq:101228a}
\end{eqnarray}
The types involved in this formula can be traced as follows:
take $id_{m\times n}$ 
and devectorize it,
obtaining $\larrow{m\times(m\times n)}{\epsilon_m}{n}$.
Then devectorize this again, getting
$\longlarrow{(n\times m)\times(m\times n)}{\unvectk n{\epsilon_m}}{1}$.
Finally, vectorize this with the product of the two thinning factors $m$ and $n$, to obtain
$\verylonglarrow{m\times n}{K_{nm} = \vectk{n\times m}{(\unvectk n{\epsilon_m})}}{n\times m}$.

The conceptual economy of (\ref{eq:101228a}) when compared with (\ref{eq:101229c})
is beyond discussion. A factorization of (\ref{eq:101228a}) can be obtained
by unfolding the $\vect{}$ and $\unvect{}$
isomorphisms:
\begin{eqnarray}
	K_{nm} &=& \vectk{n\times m}{(\unvectk n{\epsilon_m})}
	\nonumber
\\
	&=& (id_{n\times m} \otimes (\epsilon_n\comp(id_n\otimes {\epsilon_m}))) \comp \eta_{n\times m}
	\nonumber
\\
	&=& (id_{n\times m} \otimes \epsilon_n)\comp(id_{n\times m} \otimes (id_n\otimes {\epsilon_m})) \comp \eta_{n\times m}
	\nonumber
\\
	&=& (id_{n\times m} \otimes \epsilon_n)\comp
	    (id_{(n\times m)\times n} \otimes {\epsilon_m}) \comp \eta_{n\times m}
	\label{eq:101229b}
\end{eqnarray}
Listing \ref{lst:101229a} includes both versions of the commutation matrix encoded in \matlab\ notation.

Magnus and Neudecker \cite{MN79} give many properties of the commutation matrix,
including for instance,
\begin{eqnarray}
	(B\otimes A) \comp K_{ts} &=& K_{nm} \comp (A\otimes B)
	\label{eq:101228e}
\end{eqnarray}
which in our categorial setting is nothing but the statement of its \emph{naturality}
in the underlying category of matrices (polymorphism), cf.\ diagram:
\begin{eqnarray*}
\xymatrix{
	t\times s
		\ar[d]_{B\otimes A}
&
	s \times t
		\ar[l]_{K_{ts}}
		\ar[d]^{A\otimes B}
\\
	n\times m
&
	m \times n
		\ar[l]^{K_{nm}}
}
\end{eqnarray*}
Another property, not given in \cite{MN79},
\begin{eqnarray}
	K_{nn}\comp \eta_n &=& \eta_n
	\label{eq:101228d}
\end{eqnarray}
is easy to draw from  (\ref{eq:101227d}) --- just let $A := id_n$ and simplify.

\begin{lstlisting}[frame=bt,float=t,caption={Two \matlab\ encodings of
	commutation matrix
	$K_{mn}$ following (\ref{eq:101228a}) and its expansion (\ref{eq:101229b}).},label=lst:101229a]
function R = cm(n,m)
    R =  Vec(n*m,UnVec(n,epsilon(m,n)));
end

function R = cmx(n,m)
   a=kron(eye(n*m),epsilon(n,1));
   b=kron(eye((n*m)*n),epsilon(m,n));
   R = a*b*eta(n*m,m*n);
end
\end{lstlisting}

\subsection{Implementing MMM under matrix-to-vector representation} 
As we did for transpose, let us reuse previous results in
refining MMM to vectorized form. Applying (\ref{eq:101219b}) to $B$ in (\ref{eq:101124c})
we obtain, recalling type scheme (\ref{eq:110103a}):
\begin{eqnarray*}
	\vectk k{(B \comp C)} &=& (id_k \otimes (\epsilon_n\comp(id_n \otimes \vectk n B))) \comp \vectk k{C}
\end{eqnarray*}
This re-writes to
\begin{eqnarray}
	\vectk k{(B \comp C)} &=& (id_k \otimes \epsilon_n)\comp(id_{k\times n }\otimes \vectk n B) \comp \vectk k{C}
	\label{eq:110103b}
\end{eqnarray}

It may seem circular to resort to composition in the right hand side of the
above, but in fact all instances of composition there are of a special kind:
they are matrix-vector products, cf.\ linear signal transforms \cite{Voronenko:08}.
Denoting such an application of a matrix $B$ to a vector $v$ by $ap(B,v)$,
we can encode (\ref{eq:110103b}) into \matlab\ function \texttt{vecMMM} shown
in Listing \ref{lst:110103c}, under type scheme:
\begin{eqnarray*}
\xymatrix@C=8ex@R=1pt{
	j
&
	n
		\ar[l]_{B}
&
	k\
		\ar[l]_{C}
\\
	k\times j
&
	k \times (n \times j)
		\ar[l]_{id_k \otimes \epsilon_n}
&
	k \times n
		\ar[l]_{id_{k\times n} \otimes vB}
&
	1
		\ar[l]_{vC}
		\ar@/^1pc/[lll]^{vBC}
}
\end{eqnarray*}

The operator equivalent to this in the Operator Language DSL of \cite{Franchetti:09} has
interface
\begin{eqnarray*}
	\mathit{MMM}_{j,n,k}&:& \R^{jn} \times \R^{nk} \rightarrow \R^{jk}
\end{eqnarray*}
assuming the field of real numbers and vectors representing matrices in row-major order.
The operator is specified by a number of breakdown rules expressing recursive
divide-and-conquer algorithmic strategies.

\begin{lstlisting}[frame=bt,float=t,caption={\matlab\ encoding of
	vectorized MMM. Intermediate type $n$ is given explicitly.
	\texttt{vB} and \texttt{vC} are input vectors which represent composable
	matrices $B$ and $C$, respectively.
        Matrix $x$ is a constant matrix which can be made available at compile time.
%once $k$ and $j$ are inferred.
	$ap(B,v)$ denotes the application of matrix $B$ to input vector $v$.},label=lst:110103c]

function vBC = vecMMM(n,vB,vC)

    a = length(vB);
    b = length(vC);
    
    if(mod(a,n) ~= 0 || mod(b,n) ~= 0)
        error('n must be a common length factor');
    else    
        j = a / n;
        k = b / n;
        x=kron(eye(k),epsilon(n,j)); 
        y=kron(eye(k*n),vB);
        vBC = ap(x,ap(y,vC));
    end
end
\end{lstlisting}
For instance, one such rule prescribes the divide-and-conquer algorithm that splits the left-hand 
vector row-wise in a number of blocks.
Instantiating the rule to the particular case of a two-block split
corresponds to our law (\ref{eq:meither:fusion}), vectorized. The whole
OL syntax is very rich and explaining its intricacies in the current paper
would be a long detour. See Section \ref{sec:110107f} for on-going work on
typing OL formul\ae\ according to the principles advocated in the current
paper.

%
%

\section{Related Work}

Categories of matrices can be traced back to the
works of MacLane and Birkhoff \cite{MacLane98:CategoriesWM,MacLane99:Algebra},
with focus on either illustrating additive
categories or establishing a relationship between linear transformations and matrices.
Biproducts have been extensively studied in algebra and category theory. In
\cite{MacLane99:Algebra}, the same authors find applications of biproducts to
the study of additive Abelian groups and modules.
A relationship between biproducts and matrices can also be found in \cite{MacLane99:Algebra},
but it is nevertheless in \cite{MacLane98:CategoriesWM} that the hint which
triggered the current paper can be found explicit (recall Section \ref{sec:chasing}).
However, no effort on exploiting
biproducts calculationally is present, let alone algorithm derivation.
To the best of the authors' knowledge, the current paper presents the first effort to
put biproducts in the place they deserve in matrix algebra.

Bloom et al \cite{BSW96} define a generic notion of \emph{machine} and give their
semantics in terms of categories of matrices, under special (blocked) composition schemes.
They make implicit use of what we have identified as the standard
biproduct (enabling blocked matrix algebra) to formalize column and row-wise matrix join and fusion,
but the emphasis is on iteration theories which matricial theories are a particular case of.

Other categorial approaches to linear algebra include relative monads
\cite{ACU10}, whereby the category of finite-dimensional vector spaces
arises as a kind of Kleisli category.
Efforts by the mathematics of program construction community in the derivation
of matrix algorithms include the study of two-dimensional pattern
matching \cite{Je91}.

Reference \cite{BSW96} is related to Kleene algebras of matrices \cite{Ko97}.
An account of the work on calculational, index-free reasoning about
regular and Kleene algebras of matrices can be found in \cite{Ba04a}.
The close relationship between categories of matrices and relations
is implicit in the allegorial setting of Freyd and {\v{S}\v{c}edrov} \cite{FS90}:
essentially, matrices whose data values are taken from \emph{locales}
(eg.\ the Boolean algebra of truth values) are the morphisms of the corresponding allegory (eg.\
that of binary relations). Bird and de Moor \cite{BM97} follow \cite{FS90}.
Schmidt \cite{Sc10} dwells on the same relation-matrix binomial
relationship too, but from a different, set-theoretical angle. He nevertheless
pushes it quite far, eg.\ by developing a theory of vectorization
in relation algebra. Relational biproducts play no explicit role in either \cite{Sc10}, \cite{FS90}
or \cite{BM97}.

\section{Conclusions}

In this paper we have exploited the formalization of matrices as categorial
morphisms (arrows) in a way which relates categories of matrices to relation
algebra and program calculation.  Matrix multiplication is dealt with in detail, in
an elegant, calculational style whereby its divide-and-conquer, triple-nested-loop
and vectorized implementations are derived. 

The notion of a categorial biproduct is at the heart of the whole approach.
Using categories of matrices and their biproducts we have developed the algebra of
matrix-block operations and shown how biproducts scale up so as to be fit for
particular applications of linear algebra such as Gaussian elimination, for instance.

We have also shown how matrix-categorial biproducts shed light
into the essence of an important data transformation --- vectorization ---
indispensable to the efficient implementation of linear algebra packages
in parallel machines. Our calculations in this respect have shown how polymorphic
standard matrices such as eg.\ the \emph{commutation matrix} are, making dimension
polymorphism an essential part of the game, far beyond the loose
\emph{``valid only for matrices of the same order''} \cite{AM05} attitude found in the
literature. We have prototyped our constructs and diagrams in \matlab\texttrademark\ all
the way through, and this indeed showed how tedious and error-prone it is
to keep track of matrix dimensions in complex expressions. It would be much nicer to write eg.\
\texttt{eye} instead of \texttt{eye(n)}, for some hand-computed $n$ and let
\matlab\ infer which $n$ accommodate the formula we are writing.

The prospect of building biproduct-based type checkers for computer algebra
systems such as \matlab\ is therefore within reach. This seems to be 
already the approach in Cryptol \cite{LM03},
a Haskell based DSL for cryptography, where array dimensions 
are inferred using a strong type-system based on Hindley-Milner style polymorphism 
extended with arithmetic size constraints.

In retrospect, we believe to have contributed to a better understanding of
the blocked nature of linear algebra notation, which is perhaps its main advantage
--- the \emph{tremendous improvement in concision} which Backhouse stresses
in the quotation which opens the paper --- and which can be further extended
thanks to the (still to be exploited) algebra of biproducts. This raises
the issue of matrix polymorphism and enriches
our understanding that matrix dimensions are more than just numbers: they are types
in the whole sense of the word. Thus the matrix concept spruces up, raising from
the untyped number-container view (``rectangles of numbers'')
to the typed hom-set view in a category.
Perhaps Sir Arthur Eddington (1882-1944) was missing this richer view when he wrote~\footnote{
The authors are indebted to Jeremy Gibbons for pointing their attention to
this 
interesting remark of the great physicist.
}:
\begin{quote}\small
	\emph{I cannot believe that anything so ugly as multiplication
	of matrices is an essential part of the scheme of nature}
	\cite[page 39]{Ed36}.
\end{quote}

\section{Future Work} \label{sec:110107f} 

A comprehensive calculational approach to linear algebra algorithm
specification, transformation and generation is still missing. However, the
successes reported by the engineering field in automatic library generation
are a good cue to the feasibility of such a research plan. We intend to
contribute to this field of research in several directions.

\paragraph{SPIRAL} The background of our project is the formalization of
OL, the \emph{Operator Language} of \cite{Pueschel:05,Franchetti:09}, in matrix-categorial
biproduct terms. In the current paper we have stepped forward in this
direction (as compared to \cite{MO10a}, for instance) in developing a categorial
approach to vectorization, but much more work is still
needed to achieve a complete account of the refinement steps implicit in
all OL-operator breakdown rules.
SPIRAL's row-major vectorized representation calls for further work in adapting our results to such
a variant of vectorization.

\paragraph{Kleene algebras of matrices} Thus far we have assumed matrices to take their elements from
an algebraic field $K$. 
The matrix concept, however, extends to other, less rich algebraic structures,
typically involving semirings instead of
rings, for instance. Fascinating work in this wider setting shows how,
by Kleene algebra, some graph algorithms are unified into Gaussian elimination,
for instance \cite{Ba04a}. We would thus like to study the impact of such a relaxation on
our biproduct approach to the same algorithm.


\paragraph{Khatri-Rao product generalization of relational forks}
The monoidal structure provided by the tensor product defined in 
Section \ref{sec:110107d} is the key concept to generalize,
to arbitrary matrices, the relational (direct) product presented in \cite{Sc10,Fr02}.

It turns out that the \emph{fork} operation of relation algebras \cite{Fr02}
is nothing but the operator known in the linear algebra community as the Khatri-Rao
product \cite{LT05}. The standard definition offers this product as a (column-wise)
variant of Kronecker product. To emphasise the connection to relation algebra,
our definition is closer to that of a fork \cite{Fr02}: given matrices $\larrow
n A m$ and $\larrow n B k$, the Khatri-Rao product (fork) of $A$ and $B$,
denoted $\fork A B$, is the matrix of type $\larrow {n}{}{m \times k}$ defined
by
\begin{eqnarray*}
	\fork A B = (\conv\fst \comp A) * (\conv\snd \comp B)
\end{eqnarray*}
where $A * B$ is the Hadamard (element-wise) product and matrices
$\larrow{m\times k}\fst m$ and $\larrow{m\times k}\snd k$ are known
as \emph{projections}. To define these we rely on row vectors wholly filled
up with 1s, denoted by symbol ``$\bang$'' \footnote{
Notation ``$\bang$'' is imported from the algebra of programing \cite{BM97}.
}:
\begin{eqnarray*}
	\fst &=& id \otimes \bang
\\
	\snd &=& \bang \otimes id
\end{eqnarray*}

Khatri-Rao product is  associative and its unit is $\bang$, that is,
\( 
	\bang  \kr  A \wider= A \wider= A  \kr  \bang
\) hold. 
The close link between the Khatri-Rao and Kronecker products can be appreciated 
by expressing the latter in terms of the former,
\(
	A \otimes B = \fork {(A \comp \fst)}{(B  \comp \snd)}
\),
that is,
\begin{eqnarray*}
	A \otimes B = (\conv \fst \comp A \comp \fst) * (\conv \snd \comp B \comp \snd)
\end{eqnarray*}
meaning that Khatri-Rao can be used as alternative to Kronecker in formulating
concepts such as, for instance, vectorization \cite{Sc10}. A thorough comparison of
both approaches in the setting of arbitrary matrices is a topic for future work \cite{Ma11}.

\paragraph{Self-adjunctions}
The self-adjunction which supports our approach to vectorization offers a
\emph{monad} which we have not yet exploited. Already in (\ref{eq:adj:unvec:univ})
we see the unit $\larrow{n}\eta{k^2\times n}$ at work, for functor
${\fun T}\ n = k^2\times n$, whose multiplication is
of type $\larrow{k^4\times n}{\mu}{k^2\times n}$ and
can be computed following the standard theory:
\begin{eqnarray*}
	\mu & = & id \otimes \unvect{id} \wider= id \otimes \epsilon
\end{eqnarray*}
Curiously enough, the monadic flavour of vectorization can already be
savored in version (\ref{eq:110103b}) of MMM, suggesting such an implementation 
as analogue to composition in the ``brother'' Kleisli category:
\begin{eqnarray*}
	B \mathbin\bullet A &=&
	\underbrace{(id_k \otimes \epsilon_n)}_{\mu}\comp \underbrace{(id_{k\times n }\otimes B)}_{\fun F\ B} \comp A
\end{eqnarray*}
This should be studied in detail, in particular concerning the extent to
which known laws of vectorization are covered by the generic theory of monads,
discharging the corresponding proof obligations. The relationship between
this monadic setting and that of \emph{relative monads} presented in
\cite{ACU10} is another stimulus for further work in this research thread.


\subsubsection*{Acknowledgements.}
The authors would like to thank Markus P\"uschel (CMU) for driving their
attention to the relationship between linear algebra and program transformation.
 Hugo Macedo further thanks the SPIRAL group for granting him an internship
at CMU.

Thanks are also due to Michael Johnson and Robert Rosebrugh (Macquarie
Univ.) for pointing the authors to the categories of matrices approach.  Yoshiki
Kinoshita (AIST, Japan) and Manuela Sobral (Coimbra Univ.) helped with further
indications in the field.

This work is funded by ERDF - European Regional Development Fund through
the COMPETE Programme (operational programme for competitiveness) and by
National Funds through the FCT - Funda\c{c}\~ao para a Ci\^encia e a Tecnologia (Portuguese
Foundation for Science and Technology) within project FCOMP-01-0124-FEDER-010047.
Hugo Macedo holds FCT grant number {SFRH/BD/33235/2007}.

\bibliographystyle{elsarticle-num}
\bibliography{hmacedo}

\appendix

\section{Calculational proofs postponed from main text}

\paragraph{Calculation of (\ref{eq:111006a}, \ref{eq:091211a})}
The derivation of these facts is based on the existence of additive inverses
and can be found in \cite{MacLane98:CategoriesWM}. Let us see that of
(\ref{eq:111006a}) as example:
\begin{eqnarray*}
&&
	\pi_1 \comp i_2  = 0
\just\equiv{ additive inverses }
	\pi_1 \comp i_2  = \pi_1 \comp i_2 - \pi_1 \comp i_2
\just\equiv{ additive inverses }
	\pi_1 \comp i_2 + \pi_1 \comp i_2 = \pi_1 \comp i_2
\just\equiv{ (\ref{eq:biprod:a}, \ref{eq:biprod:b}) ; bilinearity (\ref{eq:090403b}, \ref{eq:091205a}) 
}
	\pi_1 \comp (i_1 \comp \pi_1 + i_2 \comp \pi_2) \comp i_2 = \pi_1 \comp i_2
\just\equiv{ (\ref{eq:biprod:c}) }
	\pi_1 \comp id \comp i_2 = \pi_1 \comp i_2
\just\equiv{ identity (\ref{eq:natid}) }
	\pi_1 \comp i_2 = \pi_1 \comp i_2
\end{eqnarray*}
The other case follows the same line of reasoning. When additive inverses
are not ensured, as in the relation algebra case, biproducts
enjoying orthogonal properties (\ref{eq:111006a},\ref{eq:091211a}) are the
ones built on top of disjoint unions in 
distributive allegories \cite{FS90,Wi00}. 

\paragraph{Proof of Theorem \ref{th:111008b}}
The calculation of (\ref{eq:biprod:a}) for biproduct (\ref{eq:101221b})
is immediate:
\begin{eqnarray*}
	\p1^C \comp i_1^C &=& \meither 1 0 \comp \msplit 1 {-C}
\\	& = & 
	\meither{id_m} 0 \comp \msplit{id_m}{-C}
\\	& = & 
	id_m
\end{eqnarray*}
The calculation of (\ref{eq:biprod:b}) is similar.
That of (\ref{eq:biprod:c}) follows:
\begin{eqnarray*}
&&
	i_1^C \comp \pi_1^C + i_2^C \comp \pi_2^C
\just={ (\ref{eq:101221b}) }
	\msplit 1 {-C} \comp \meither 1 0 + \msplit 0 1 \comp \meither {C} 1
\just={ fusion laws (\ref{eq:meither:fusion}, \ref{eq:msplit:fusion}) twice }
	\mblock 1 0 {-C} 0 + \mblock 0 0 {C} 1
\just={ blocked addition (\ref{eq:101221c}) }
\eqnnewpage
	\mblock 1 0 0 1
\just={ standard biproduct reflection (\ref{eq:091028a},\ref{eq:091028b}) }
	id
\end{eqnarray*}

\paragraph{Proof of Theorem \ref{th:111008c}}
Only the calculation of (\ref{eq:biprod:c}) is given below,
those of (\ref{eq:biprod:a}) and (\ref{eq:biprod:b}) being similar
and actually simpler. Dropping identity matrix subscripts and relying on composition
binding tighter than $\otimes$ to save parentheses, we reason:
\begin{eqnarray*}
&&
	(i_1 \otimes id) \comp (\pi_1 \otimes id) + (i_2 \otimes id) \comp (\pi_2 \otimes id)
\just={ (\ref{eq:tensor:functor}) twice ; (\ref{eq:natid}) twice }
	(i_1 \comp \pi_1 \otimes id) + (i_2 \comp \pi_2 \otimes id)
\just={ (\ref{eq:101218e}) }
	(i_1 \comp \pi_1 + i_2 \comp \pi_2) \otimes id
\just={ (\ref{eq:biprod:c})  }
	id \otimes id
\just={ (\ref{eq:101218c}) }
	id
\end{eqnarray*}

\paragraph{Calculation of fusion law (\ref{eq:101218g})}
\begin{eqnarray*}
&&
	\meither A B \otimes C
\just={ (\ref{eq:meither:def}) }
	(A \comp \pi_1 + B \comp \pi_2) \otimes C
\just={ (\ref{eq:101218e}) }
	(A \comp \pi_1 \otimes C) + (B \comp \pi_2 \otimes C)
\just={ $C = C \comp id$ twice ; (\ref{eq:tensor:functor}) twice }
	(A \otimes C) \comp (\pi_1 \otimes id) +
	(B \otimes C) \comp (\pi_2 \otimes id)
\just={ (\ref{eq:101218f}) }
	(A \otimes C) \comp \pi_1 + (B \otimes C) \comp \pi_2
\just={ (\ref{eq:meither:def}) }
	\meither{A \otimes C}{B \otimes C}
\end{eqnarray*}
The elegance of this calculation compares favourably with the telegram-like
proof of a similar result in \cite{AM05} (``Kronecker product of a partitioned
matrix'') carried out at index-level, using ``dot-dot-dot'' notation.

\paragraph{Calculation of (\ref{eq:101218fh})}
\begin{eqnarray*}
&&
	\epsilon \otimes id
\just={ (\ref{eq:101221i}) }
	\meither{\p1}{\p2} \otimes id
\just={ (\ref{eq:101218g}) }
	\meither{\p1 \otimes id}{\p2 \otimes id}
\just={ (\ref{eq:101218f}) twice; (\ref{eq:101221i}) }
	\epsilon
%
%
\end{eqnarray*}

\paragraph{Calculation of (\ref{eq:vecOfSum})}
\begin{eqnarray*}
&&
	\vect(A+B) =  \vect A + \vect B
\just\equiv{ universal property (\ref{eq:101124a})  }
	A+B = \epsilon \comp (id \otimes (\vect A+\vect B))
\just\equiv{ Kronecker product (\ref{eq:101218d}) }
	A+B = \epsilon \comp (id \otimes \vect A + id\otimes\vect B)
\just\equiv{ composition is bilinear (\ref{eq:090403b}) }
	A+B = \epsilon \comp (id \otimes \vect A) + \epsilon \comp (id\otimes\vect B)
\just\equiv{ cancellation law (\ref{eq:101219b})  twice }
	A+B = A+B
%
%
\end{eqnarray*}

\paragraph{Calculation of (\ref{eq:101124c})}
This follows by instantiating cancellation law (\ref{eq:adj:unvec:canc}),
for $A := \vect{(B \comp C)}$,
knowing that $\vect\relax$ and $\unvect\relax$ are inverses:
\begin{eqnarray*}
&&
	\vect{(B \comp C)}
\just={(\ref{eq:adj:unvec:canc})}
	(id  \otimes (B \comp C)) \comp \eta
\just={ identity (\ref{eq:natid}) }
	(id \comp id) \otimes (B \comp C) \comp \eta	
\just={ bifunctoriality (\ref{eq:tensor:functor}) ; associativity (\ref{eq:comp}) }
	(id \otimes B) \comp ((id \otimes C) \comp \eta)
\just={ canceling (\ref{eq:adj:unvec:canc}) }
	(id \otimes B) \comp \vect{C}
\end{eqnarray*}

\paragraph{Calculation of (\ref{eq:101124d})}
We reason, minding subscripts $k$, $m$ and $n$:
\begin{eqnarray*}
&&
	\vectk m(C\comp B) = (\conv B \otimes id_n) \comp \vectk k C
\just\equiv{ (\ref{eq:101224c}) twice }
	(id_m \otimes (C\comp B)) \comp \eta_m= (\conv B \otimes id_n) \comp (id_k \otimes C) \comp \eta_k
\just\equiv{ (\ref{eq:tensor:functor}) twice; (\ref{eq:natid}) }
	(id_m \otimes C) \comp (id_m \otimes B) \comp \eta_m =
	(id_m \otimes C) \comp (\conv B \otimes id_k)  \comp \eta_k
\just\implied{ Leibniz }
	(id_m \otimes B) \comp \eta_m= (\conv B \otimes id_k)  \comp \eta_k
\just\equiv{ see (\ref{eq:101219a}) below}
	true
\end{eqnarray*}
The calculation relies on the commutativity of diagram
\begin{eqnarray}
\xymatrix@C+2em{
	m
		\ar[d]_{B}
&
	m^2
		\ar[d]_{id_m    \otimes B   }
&
	1
		\ar[dl]_{\vectk m B}
		\ar[l]_{\eta_m}
		\ar[d]^{\eta_k}
\\
	k
&
	m \times k
&
	k^2
		\ar[l]^{\conv B \otimes id_k}
}
	\label{eq:101219a}
\end{eqnarray}
whose proof amounts to justifying equation
\begin{eqnarray}
	\vectk m B & = & (\conv B \otimes id_k) \comp \eta_k
	\label{eq:101227b}
\end{eqnarray}
for $\rarrow{m}{B}{k}$. Changing variable $A :=\conv B$,
\begin{eqnarray}
	\vectk m{(\conv A)}  & = & (A \otimes id_k) \comp \eta_k
	\label{eq:101228b}
\end{eqnarray}
we see that it means that, by swapping the terms of the Kronecker product in a vectorization
of a matrix $\rarrow{k}{A}{m}$, we produce a row-major vectorization of $A$ instead of column-major one.
This is amply discussed in \cite{Johnson:90}. 

The calculation of (\ref{eq:101228b}) relies on known properties of the commutation matrix:
\begin{eqnarray*}
	\vectk m{(\conv A)} &=& K_{mk} \comp \vectk k A
\just\equiv{ (\ref{eq:101224c}) }
	K_{mk} \comp (id_k \otimes A) \comp \eta_k
\just\equiv{ natural-$K$ (\ref{eq:101228e}) }
	(A \otimes id_k) \comp K_{kk} \comp \eta_k
\just\equiv{ (\ref{eq:101228d}) }
	(A \otimes id_k) \comp \eta_k
\end{eqnarray*}

\noindent
Calculation of (\ref{eq:110103d}):
\begin{eqnarray*}
&&
	\vectk {k+k'}{\meither A B} 
\just={ either def. (\ref{eq:meither:def})}
	\vectk {k+k'}{(A \comp \p1 + B \comp \p2)} 
\just={ linearity of $\vect$ (\ref{eq:vecOfSum})}
	\vectk{k+k'}{(A \comp \p1)} + \vectk {k+k'}{(B \comp \p2)} 
\just={ $\vect$ of composition (\ref{eq:101124c})}
	(id_{k+k'} \otimes A) \comp (\vectk{k+k'}{\p1}) + (id_{k+k'} \otimes B) \comp (\vectk {k+k'}{\p2})
\just={ $\vect$ of composition (\ref{eq:101124c}) ($\p1 = id_k \comp \p1$) and  ($\p2 = id_{k'} \comp \p2$)  }
	(id_{k+k'} \otimes A) \comp ((\conv{\p1} \otimes id_k) \comp \vectk{k}{id}) + (id_{k+k'} \otimes B) \comp ((\conv{\p2} \otimes id_{k'}) \comp \vectk{k'}{id})
\just={ duality (\ref{eq:091206a}) ; scaled injections (\ref{eq:111008d}) ; $\eta$ def. (\ref{eq:111008e}) }
	(id_{k+k'} \otimes A) \comp i_1 \comp \eta_k + (id_{k+k'} \otimes B) \comp i_2 \comp \eta_{k'} 
\just={ functor bilinearity (\ref{eq:101218e})  $(id_{k+k'} = id_k \oplus id_{k'})$ }
%
%
	((id_{k} \otimes A) \oplus (id_{k'} \otimes A)) \comp i_1 \comp \eta_k + ((id_{k} \otimes B) \oplus id_{k'} \otimes B)) \comp i_2 \comp \eta_{k'} 
\just={ naturality (\ref{eq:injnatprop}) }
	i_1 \comp (id_{k} \otimes A) \comp \eta_k + i_2 \comp (id_{k'} \otimes B) \comp \eta_{k'} 
\just={ $\vect$ definition (\ref{eq:101224c}) }
	i_1 \comp \vectk{k}{A} + i_2 \comp \vectk{k'}{B} 
\just={ split definition (\ref{eq:msplit:def}) }
	\msplit{ \vectk{k}{A}}{\vectk{k'}{B}} 
\end{eqnarray*}

\end{document}